\documentclass[12pt]{article}

\usepackage{amsmath}
\usepackage{amssymb}
\newcommand{\al}{\alpha}
\newcommand{\ep}{\epsilon}

\newcommand{\lb}{\lbrack}
\newcommand{\rb}{\rbrack}

\newcommand{\msc}[1]{\mbox{\scriptsize #1}}
\newcommand{\dsp}{\displaystyle}

\newcommand{\bc}{\Bbb C}
\newcommand{\br}{\Bbb R}
\newcommand{\bz}{\Bbb Z}

\newcommand{\bh}{\Bbb H}

\newcommand{\bsz}{\Bbb Z}

\newcommand{\g}{\mbox{{\bf g}}}
\newcommand{\f}{\mbox{{\bf f}}}

\renewcommand{\-}{{\bf -1}}

\newcommand{\cA}{{\cal A}}

\newcommand{\cJ}{{\cal J}}
\newcommand{\cT}{{\cal T}}

\newcommand{\cN}{{\cal N}}
\newcommand{\cM}{{\cal M}}

\newcommand{\cP}{{\cal P}}

\newcommand{\cC}{{\cal C}}

\newcommand{\cD}{{\cal D}}

\newcommand{\cH}{{\cal H}}
\newcommand{\cU}{{\cal U}}

\newcommand{\cI}{{\cal I}}

\newcommand{\tL}{\tilde{L}}
\newcommand{\tJ}{\tilde{J}}
\newcommand{\tI}{\tilde{I}}
\newcommand{\tG}{\tilde{G}}

\newcommand{\tcI}{\widetilde{\cal I}}

\newcommand{\tchi}{\tilde{\chi}}
\newcommand{\tell}{\tilde{\ell}}

\newcommand{\hsigma}{\widehat{\sigma}}

\newcommand{\ket}[1]{{\left|#1\right\rangle}}

\newcommand{\dket}[1]{{\left.\left|#1\right\rangle\right\rangle}}
\newcommand{\dbra}[1]{{\left\langle\left\langle#1\right|\right.}}

\newcommand{\Th}[2]{\Theta_{#1,#2}}
\renewcommand{\th}{{\theta}}

\newcommand{\ch}[2]{\mbox{ch}^{#1}_{#2}}

\newcommand{\tr}{\mbox{Tr}}
\renewcommand{\mod}{\mbox{mod}}

\newcommand{\nn}{\nonumber\\}

\newcommand{\NS}{\mbox{NS}}
\newcommand{\tNS}{\widetilde{\mbox{NS}}}
\newcommand{\R}{\mbox{R}}
\newcommand{\tR}{\widetilde{\mbox{R}}}
\newcommand{\sNS}{\msc{NS}}
\newcommand{\stNS}{\widetilde{\msc{NS}}}
\newcommand{\sR}{\msc{R}}
\newcommand{\stR}{\widetilde{\msc{R}}}

\newcommand{\any}{{}^{\forall}}

\newcommand{\bl}{{\bf l}}
\newcommand{\tbl}{\tilde{\bl}}

\newcommand {\eqn}[1]{(\ref{#1})}

\makeatletter
\@addtoreset{equation}{section}
\def\theequation{\thesection.\arabic{equation}}
\makeatother

\usepackage{color,comment}

\begin{document}


\begin{titlepage}
 \
 \renewcommand{\thefootnote}{\fnsymbol{footnote}}
 \font\csc=cmcsc10 scaled\magstep1
 {\baselineskip=16pt
  \hfill
 \vbox{\hbox{April, 2017}
 \hbox{UTHEP-702}
       }}

 \baselineskip=20pt
\vskip 1cm
 
\begin{center}

{\bf \Large 

Non-supersymmetric D-branes 

with Vanishing Cylinder Amplitudes

in Asymmetric Orbifolds

} 

 \vskip 1.2cm

\noindent{ \large Yuji Satoh}\footnote{\sf ysatoh@het.ph.tsukuba.ac.jp}
 \\

\medskip

{\it Institute of Physics, University of Tsukuba, \\
 Ibaraki 305-8571,  Japan}

\vskip 8mm
\noindent{ \large Yuji Sugawara}\footnote{\sf ysugawa@se.ritsumei.ac.jp},
\hspace{1cm}
\noindent{ \large Takahiro Uetoko}\footnote{\sf rp0019fr@ed.ritsumei.ac.jp},

\medskip

 {\it Department of Physical Sciences, 
 College of Science and Engineering, \\ 
Ritsumeikan University,  
Shiga 525-8577, Japan}

\end{center}

\bigskip

\begin{abstract}

We study the type II string vacua with chiral space-time SUSY constructed as asymmetric orbifolds of 
torus and $K3$ compactifications. Despite the fact that all the D-branes are non-BPS 
in any chiral SUSY vacua, 
we show that the relevant non-geometric vacua of asymmetric orbifolds allow rather generally 
configurations of D-branes which lead to vanishing cylinder amplitudes, implying
the bose-fermi cancellation at each mass level of the open string spectrum.  
After working on simple models of  toroidal asymmetric orbifolds, we  focus 
on the asymmetric orbifolds of $T^2 \times \cM$, 
where $\cM$ is described by 
a general $\cN=4$ SCFT with $c=6$ defined by the Gepner construction for $K3$.
Even when the modular invariant partition functions 
in the bulk
remain  unchanged, 
the spectra of such non-BPS D-branes with the bose-fermi cancellation 
can vary significantly according to the choice of orbifolding.

\end{abstract}

\setcounter{footnote}{0}
\renewcommand{\thefootnote}{\arabic{footnote}}

\end{titlepage}

\baselineskip 18pt

\vskip2cm 
\newpage


\section{Introduction}

String theories on the {\em non-geometric\/} backgrounds may 
induce interesting features which are 
not  realized in the standard geometric compactifications.  
One of the salient aspects of such 
non-geometric vacua would be the vanishing cosmological constant without unbroken SUSY.
This is in contrast to our experiences in ordinary geometric string vacua that 
the SUSY-violation generically gives rise to  cosmological constant at the breaking mass scale 
(string scale, typically).
The attempts of the construction of non-SUSY vacua with vanishing cosmological constant 
have been initiated by  \cite{Kachru1,Kachru2,Kachru3} 
based on some non-abelian orbifolds, followed by closely related studies {\em e.g.} in 
\cite{Harvey,Shiu-Tye,Blumenhagen:1998uf,Angelantonj:1999gm,Antoniadis,Aoki:2003sy}. 
More recently, several non-SUSY vacua with this property have been  constructed  
as asymmetric orbifolds \cite{Narain:1986qm} 
by simpler cyclic groups 
in \cite{SSW,SWada}.
Studies of non-SUSY vacua in heterotic string theory have 
also been presented 
{\em e.g.} in
\cite{Blaszczyk:2014qoa,Angelantonj:2014dia,Faraggi:2014eoa,Abel:2015oxa,
Kounnas:2015yrc,Abel:2017rch}.


In this paper, we would like to focus on similar interesting aspects of non-BPS D-branes 
in simple models of non-geometric type II string vacua. 
Let us first recall that 
the BPS D-branes are described by the boundary states satisfying the BPS-equation,
\begin{equation}
\left[ Q^{\al} + M^{\al}_{~\beta} \widetilde{Q}^{\beta} \right] \, \dket{B} =0, 
\label{BPS cond}
\end{equation}
where $Q^{\al}$ ($\widetilde{Q}^{\beta}$) denotes the left(right)-moving space-time supercharges 
and $M^{\al}_{~\beta}$ are some c-number coefficients.
Through this paper, we express  boundary states by $\dket{\cdots }$, $\dbra{\cdots}$.
We then anticipate that the cylinder amplitude of which both ends are attached to the common 
BPS D-brane, which we  call the
`self-overlap' in this paper,  should vanish, 
\begin{equation}
Z_{\msc{cyl}}(s) \equiv \dbra{B} e^{-\pi s H^{\msc{(c)}} } \dket{B} =0.
\label{vanishing cyl}
\end{equation}
Here we identify $s \in \br_{>0}$ as the closed string modulus and  $t\equiv 1/s$  as the open string one. 
Needless to say, this means that we have a precise bose-fermi cancellation at each mass level 
in the open string spectrum, naturally expected from the BPS property of 
the D-brane. 
However, the bose-fermi cancellation \eqn{vanishing cyl} does not necessarily imply 
that the boundary state $\dket{B}$ satisfies the BPS equation \eqn{BPS cond}.
Indeed, it has been known that, in some  superstring vacua, 
there exist non-BPS configurations of D-branes that however realize the bose-fermi cancellation 
of open strings \cite{Blumenhagen:1998uf,GSen}.
The main purpose of this paper is to demonstrate 
that {\em non-geometric\/} backgrounds of superstring theory  rather generally 
accommodate such non-BPS D-branes with vanishing cylinder amplitudes.

Although we concentrate in this paper mainly on the theoretical aspects from
the view points of world-sheet conformal field theory, 
we would also like to mention  
a `physical' motivation of this work: 
Since the closed string sector in the bulk is supersymmetric in our setting,
the supersymmetry would be broken solely by the effect of the non-BPS D-branes.
More concretely,  
if we have sufficiently generic configurations of the non-BPS  D-branes as above, 
the SUSY-breaking would be brought about by the condensation of the 
non-BPS `D-brane instantons' (Euclidean D-branes wrapping 
around internal cycles).
In such a case,
 because  the $O(g_s^0)$-contributions to the cosmological constant, 
 as well as the bulk ones,  still vanish due to \eqn{vanishing cyl},  
we would be left with a non-perturbatively small cosmological constant induced by the instanton effect, 
which is exponentially suppressed as long as the string coupling $g_s$ is sufficiently small.
Such a possibility in a  type II theory has indeed been mentioned in \cite{Harvey}  based on the analysis 
of its heterotic dual.
The present work may be a step toward realizing such string vacua with small cosmological constant.


Now, let us make a brief sketch of our basic idea: 
\begin{itemize}
\item We start with the type II superstring vacua preserving only 
the {\em chiral\/} SUSY,  
which are straightforwardly constructed by the asymmetric orbifolding 
by the twist $\sigma$ 
that eliminates, say, all the left-moving supercharges $Q^{\al}$.

\item In these vacua, while the cosmological constant in the bulk should vanish 
due to the existence of unbroken SUSY, 
{\em any\/} D-branes cannot be BPS. 
In other words, any boundary states cannot satisfy the BPS equation \eqn{BPS cond}  
due to the lack of $Q^{\al}$.

\item We search for the boundary states realizing nevertheless the vanishing self-overlap \eqn{vanishing cyl},
 which are obtained from the BPS D-branes $ \dket{B}_0$ in the untwisted theory 
 by the orbifold projection, 
$\dket{B} \propto {\cal P} \dket{B}_0$.
The conformal invariance is maintained, since $ {\cal P}$ commutes with the Virasoro operators.

\end{itemize}


Of course, in generic chiral SUSY vacua, there are no solutions of the boundary states with 
the vanishing self-overlaps. 
However, once the asymmetric twist 
to preserve the chiral SUSY is given, 
the self-overlap of the projected D-brane $ \dket{B}$ is likely to be vanishing as long as
it inherits the structure of the bose-fermi cancellation in the bulk torus amplitude.
As shown in the following sections,
it is indeed possible to find simple models of such asymmetric orbifolds, 
and thus plenty of boundary states with the vanishing self-overlap.
In section 2, we study toroidal models and consider several asymmetric orbifoldings
 preserving 8 supercharges coming only from the right-mover.

In section 3, which is the main part of this paper,
we shall discuss less supersymmetric 
models constructed as the asymmetric orbifolds of the backgrounds, 
\begin{equation}
\br^{3,1}\times T^2 \times \cM,
\label{background 0}
\end{equation}
where $\cM$ is described by 
a general $\cN=4$ superconformal field theory (SCFT) 
with $\hat{c} (\equiv \frac{c}{3})=2$,
which geometrically describes compactifications on $K3$ with particular moduli. 
The relevant asymmetric orbifolds are defined by the twisting, 
\begin{equation}
\sigma = (\-_R)^{\otimes 2} \otimes \sigma_{\cM},
\label{sigma 0}
\end{equation}
where $(\-_R)^{\otimes 2}$ is the chiral reflection on the $T^2$-sector ($X^{4,5}$-directions), 
\begin{equation}
(\-_R)^{\otimes 2}\,:\, (X^{i}_L, X^{i}_R) ~ \longmapsto ~  ( X^{i}_L, - X^{i}_R), 
~~~  
 (\psi^{i}_L, \psi^{i}_R) ~ \longmapsto ~  (\psi^{i}_L, - \psi^{i}_R), 
\hspace{1cm} (i=4,5),
\end{equation}
and $\sigma_{\cM}$ denotes an involution on the $\cM$-sector, which is allowed to act asymmetrically 
on the $\cN=4$ superconformal algebra (SCA). 
As we will clarify later, one obtains 
in this way
the chiral SUSY vacua with
the 4-dim. $\cN=1$ SUSY (4 supercharges). 
We then classify the possible gluing conditions for the boundary states, 
which are decomposed into the Ishibashi states \cite{Ishibashi} for each $\cN=4$ unitary 
irreducible representations (irrep.'s), 
and 
examine whether or not their self-overlaps vanish. 
The spectra of the non-BPS boundary states with this property  non-trivially depend on   
the choice of the twist operator $\sigma_{\cM}$, even in the cases 
when the modular invariant partition functions remain unchanged; 
different $\sigma_{\cM}$'s may lead to the same partition functions in the bulk.

~


\section{Toroidal Asymmetric Orbifolds}
\label{toroidal models}

In this section we shall focus on the simpler cases, namely, the asymmetric orbifolds of tori 
realizing the chiral SUSY vacua of type II string,
in order to show how the strategy outlined above is implemented.
The discussion is straightforwardly extended to the case of $K3$ in the next section, though 
it is technically a little more involved.

\subsection{
Asymmetric Orbifold 
\ $T^4[D_4]/\left[ (-1)^{F_L}\otimes (\-_R)^{\otimes 4}\right]$
}
\label{subsec:torus1}

Let us first consider the asymmetric orbifold of the 4-dim. tours $T^4$, which 
would be the simplest model that has the desired properties.
We assume the torus is along the $X^{6, \ldots, 9}$-directions and 
at the symmetry enhancement point with $\widehat{SO}(8)_1$.
We thus denote it as $T^4[D_4]$,
the corresponding partition function of which reads
\begin{equation}
 Z^{T^4[D_4]}(\tau,\bar{\tau}) 
= \frac{1}{2}\left\{
\left|\frac{\th_3}{\eta}\right|^8+\left|\frac{\th_4}{\eta}\right|^8
+\left|\frac{\th_2}{\eta}\right|^8
\right\}.
\label{Z D4}
\end{equation}
The orbifold group is generated by a single element 
\begin{equation}
\sigma \equiv (-1)^{F_L}\otimes (\-_R)^{\otimes 4},
\label{sigma 1}
\end{equation}
which acts as the chiral reflection on 
the right-mover, 
$X^i_R\, \rightarrow \,
-X^i_R$, $\psi_R^i\, \rightarrow \, -\psi_R^i$ ($i=6,\ldots, 9$), 
accompanied by the twisting 
of the space-time fermion number 
$(-1)^{F_L}$ on the left-moving fermions, that is, the sign-flip 
of arbitrary states in the left-moving R(amond)-sector. 
Closely related  asymmetric orbifolds adopting slightly different setting
have been analyzed in the bulk  \cite{SSW,SWada} for non-supersymmetric string vacua
with vanishing cosmological constant.
The analysis below follows  these references.

We simply assume that 
$\sigma^2$ acts on the untwisted Hilbert space
as an involution for 
the free bosons $X^i_R$, 
whereas we naturally have two possibilities on the fermionic sector;
(i) $\sigma^2= {\bf 1}$, ~ 
(ii) $\sigma^2= (-1)^{F_R}$, 
depending on the definition of the Ramond vacua
or the way of bosonization to introduce the spin fields (see also section \ref{orbifoldT2K3}).  
Here, the operator $(-1)^{F_R}$ just acts as the sign flip on any states in the right-moving R-sector.  
We separately examine theses two cases:

~


\noindent
{\bf (i) $\sigma^2= {\bf 1}$ ~(on the untwisted Hilbert space) } 

In this case, the modular invariant is written as 
\begin{align}
Z(\tau,\bar{\tau}) & =  Z_{\msc{bosonic}}^{6d}(\tau,\bar{\tau}) \, \frac{1}{4} \sum_{a,b \in \bz_4}\, 
Z^{T^4[D_4]}_{(a,b)}(\tau,\bar{\tau})
 h_{(a,b)}(\tau) \overline{f_{(a,b)}(\tau)},
\nn
Z^{T^4[D_4]}_{(a,b)}(\tau,\bar{\tau}) & := 
\left\{
\begin{array}{ll}
 Z^{T^4[D_4]}(\tau,\bar{\tau}) & ~~ (a,b \in 2\bz),
\\
\ep_{(a,b)}^{[4]} \chi^{D_4}_{(a,b)}(\tau) \overline{\left(\tchi^{A_1}_{(a,b)}(\tau)\right)^4} 
& ~~ (a \in 2\bz+1, ~ \mbox{or} ~ b \in 2\bz+1).
\end{array}
\right.
\label{Z 1-1}
\end{align}
where $Z_{\msc{bosonic}}^{6d}(\tau,\bar{\tau})$ denotes the partition
function of the bosonic sector of uncompactified space-time $\br^{5,1}$.
The building blocks $\chi^{D_4}_{(a,b)}$, $\overline{\left(\tchi^{A_1}_{(a,b)}(\tau)\right)^4}$,  
$h_{(a,b)}$ and $\overline{f_{(a,b)}}$ are 
evaluated for the sectors 
of $X_L^{6, \ldots, 9}$, $X_R^{6, \ldots, 9}$,
the left-moving fermions, and the right-moving fermions,
respectively,
where the subscript $(a,b)$ labels  the sectors with the spatial and temporal twists
by $\sigma$ given in \eqn{sigma 1}.
They are obtained first for the (0,1) sector with one temporal twist, and then for other sectors
by the modular transformation.
Their explicit forms are summarized in Appendix A. (See \eqn{Dr ab}, \eqn{tA1 ab}, \eqn{fab}, \eqn{hab}.)
The phase factor $\ep^{[r]}_{(a,b)}$ is defined in \cite{SatohS}, and  explicitly written as  
\begin{equation}
\ep^{[r]}_{(a,b)} :=  e^{\frac{i\pi}{8} r (-1)^a ab} \left(\kappa_{(a,b)}\right)^r, \hspace{1cm} (a \in 2\bz+1 ~ \mbox{or} ~ b \in 2\bz +1),
\label{ep r}
\end{equation}
with 
\begin{equation}
\kappa_{(a,b)} := 
\left\{
\begin{array}{ll}
-1 & ~~~ a \equiv 3,5 ~ (\mod \, 8), ~ b \in 2\bz+1,
\\
1 & ~~~ \mbox{otherwise}.
\end{array}
\right.
\end{equation}
It is quite useful to 
note  that the combination 
$\ep^{[r]}_{(a,b)} \chi^{X_r}_{(a,b)}(\tau) \overline{\left(\tchi^{A_1}_{(a,b)}(\tau)\right)^r} $
(or $\ep^{[-r]}_{(a,b)}  \left(\tchi^{A_1}_{(a,b)}(\tau)\right)^r \overline{\chi^{X_r}_{(a,b)}(\tau)} $)
is organized so as to be modular covariant with respect to $(a,b)$, 
where $X_r$ denotes the suitable Lie algebra lattice of rank $r$ presented in \cite{SatohS}.
Namely, any modular transformation defined by $A \in SL(2;\bz)$ acts simply on the subscript $(a,b)$ as 
$
(a,b) \, \longmapsto \, (a,b) A.
$
We note that this is an order 4 orbifold due to the existence of the phase factor \eqn{ep r}
despite  
$\sigma^2 = {\bf 1}|_{\msc{untwisted}}$, 
which would be a typical feature in asymmetric orbifolds. 

The right-mover preserves 1/2 space-time SUSY, whereas the left-moving
space-time SUSY is completely broken. In fact, it is obvious that 
$\sigma\equiv (-1)^{F_L} \otimes (\-_R)^{\otimes 4}$ cannot preserve any
left-moving supercharges in the 
even $a$ sector, which are essentially those in the unorbifolded theory. 
Furthermore, if we had a left-moving supercharges belonging to the sector $a=1$, 
we should obtain the equality of the partition functions 
\begin{equation}
Z^{(\sNS, \sNS)}_{a=0}(\tau,\bar{\tau}) = - Z^{(\sR,\sNS)}_{a=1}(\tau,\bar{\tau}).
\end{equation}
However, it is easy to see that this is not the case, when observing the explicit forms 
of relevant partition functions. 
We can similarly show the absence of supercharges in the $a=-1$ sector.
On the other hand, half of untwisted supercharges in the right-mover are 
$\sigma$-invariant, as in the familiar supersymmetric orbifold 
$T^4/\left[(\-_L)^{\otimes 4} \otimes (\-_R)^{\otimes 4}\right]$.


Now, let us move on to the discussion on the non-BPS D-branes.
As already pointed out, no D-brane can preserve space-time SUSY. 
Nevertheless, rather general `bulk-type branes' lead to the vanishing self-overlap.%
\footnote
{We shall call the boundary states made up only by the untwisted sector
as the `bulk-type' to distinguish them from the `fractional branes' that include the contributions 
from the twisted sectors.}
In fact, consider the  bulk-type brane written as 
an orbifold projection,
\begin{eqnarray}
 && 
  \dket{B} = \sqrt{2} {\cal P} \dket{B}_0, 
\label{desired brane 1}
\end{eqnarray}
where 
$\dket{B}_0$ stands for the GSO-projected 
boundary state describing any BPS
D-brane in the unorbifolded theory on $\br^{5,1}\times T^4[D_4]$,
and ${\cal P} = \frac{1}{2} (1+\sigma)$
is the projection operator
onto the invariant sector under the twist.
As described in the introduction, ${\cal P}$ commutes with the Virasoro
operators and maintains the conformal invariance.
The overall normalization factor $\sqrt{2}$ has been determined by the Cardy
condition. 
By definition, we have 
$
{}_0\dbra{B} e^{-\pi s H^{(c)}} \dket{B}_0 = 0, 
$
since $\dket{B}_0$ is BPS. 
Moreover, explicit computation gives
\begin{eqnarray}
 {}_0\dbra{B} \sigma \, e^{-\pi s H^{(c)}} \dket{B}_0
& \equiv &  {}_0\dbra{B} \left[(-1)^{F_L}\otimes (\-_R)^{\otimes 4}\right] \, e^{-\pi s H^{(c)}}
\dket{B}_0 
\propto  f_{(0,1)}(is) \equiv 0.
\label{Zcyl sigma torus 1}
\end{eqnarray}
Again $f_{(0,1)}(is)$ is defined in \eqn{fab}.
We thus obtain 
\begin{align}
& \dbra{B} e^{-\pi s H^{(c)}} \dket{B} =
{}_0\dbra{B} \, e^{-\pi s H^{(c)}} \dket{B}_0 + {}_0\dbra{B} \, \sigma e^{-\pi s H^{(c)}} \dket{B}_0 =  0.
\label{eval overlap 1}
\end{align}
Although the left- and right-movers are correlated in the boundary states due to the conformal
invariance, the twist thereon still leads to the same function $f_{(0,1)}$ as in the bulk,
which is regarded as a remnant of the bulk computation.
In this way, we have successfully shown that the present string vacuum possesses the desired property
to have the non-BPS D-branes with vanishing self-overlaps.

Because of the overall factor in $\dket{B}$, its coupling to the gravitons (tension) is 
$\sqrt{2}$ times that in the unorbifolded theory. 
The coupling to the RR-particles (RR charge) is also multiplied by $\sqrt{2}$.
By the modular transformation, the standard open string excitations in the original theory 
are found to remain in the self-overlap of the unorbifolded part $\dket{B}_0$.
These are common features for all the non-BPS branes with the vanishing self-overlaps 
treated in this paper. 

~


\noindent
{\bf Absence of tachyonic instability} 

Let us briefly check that no open string tachyons emerge in 
the cylinder amplitude,
$$
Z_{\msc{cylinder}}(it) = \dbra{B} e^{-\pi s H^{(c)}} \dket{B}, \hspace{1cm} 
\left(t \equiv 1/s\right).
$$
In fact, the piece ${}_0\dbra{B}  e^{-\pi s H^{(c)}} \dket{B}_0$ 
is just the same as the  familiar cylinder amplitude associated to the BPS brane, 
whereas
\begin{eqnarray}
{}_0\dbra{B} \sigma \, e^{-\pi s H^{(c)}} \dket{B}_0
&\propto& 
\left(\sqrt{\frac{2 \eta(is)}{\th_2(is)}}\right)^4  \cdot f_{(0,1)}(is) \equiv 
\frac{\th_3(is)^2 \th_4(is)^2}{\eta(is)^4} \cdot f_{(0,1)}(is)
\nn
&= & 
\frac{\th_3(it)^2 \th_2(it)^2}{\eta(it)^4}\cdot f_{(1,0)}(it)
\equiv \frac{\th_3(it)^4 \th_2(it)^4}{2\eta(it)^8} - \frac{\th_2(it)^4 \th_3(it)^4 }{2\eta(it)^8}.
\end{eqnarray}
In the last line, the first and second terms are identified as the NS and R-sector amplitudes in the open 
string channel, of which leading terms are obviously massless. 
We then obtain 16 pairs of massless bosonic and fermionic states from the orbifolded part,
even though no supercharges in the closed string sector preserve the boundary state $\dket{B}$.  
We can similarly show the absence of open string tachyons in the cylinder amplitudes with the bose-fermi cancellation also for other models discussed below.

~


\noindent
{\bf (ii)  $\sigma^2= (-1)^{F_R} $ ~(on the untwisted Hilbert space)} 

In this case, $\sigma$ acts as a $\bz_4$-action already on the untwisted sector, and 
the modular invariant is slightly modified as 
\begin{eqnarray}
 && Z(\tau,\bar{\tau}) = Z_{\msc{bosonic}}^{6d}(\tau,\bar{\tau}) \, \frac{1}{4}
\sum_{a,b\in \bsz_4}\, 
Z^{T^4[D_4]}_{(a,b)}(\tau,\bar{\tau}) h_{(a,b)}(\tau)\overline{\f_{(a,b)}(\tau)}.
\label{Z 1-3}
\end{eqnarray}
The fermionic chiral block $h_{(a,b)}$ is again given in  \eqn{hab}, 
while $\f_{(a,b)}$, given in \eqn{fab2}, is slightly modified from $f_{(a,b)}$ 
due to the relation $\sigma^2= (-1)^{F_R} $.
The left-mover has no space-time SUSY as in the first model.
At first glance, it seems that the right-moving SUSY is also broken, 
because all of the supercharges in the untwisted sector are projected out by $(-1)^{F_R}$. 
However, it is found that (NS,R)-massless states appear in the $a=2$ twisted sector,  
suggesting the existence of new 8 supercharges. 
These states possess the opposite chirality to the case (i), because the orbifolding by $(-1)^{F_R}$ 
acts like the T-duality transformation (see {\em e.g.} \cite{GSen,Sen}).
In the end, we indeed obtain a chiral SUSY vacuum. 
One can check that the partition function vanishes after summing up $a, b \in 2 \bz$, 
although each  $\overline{\f_{(a,b)}(\tau)}$  is not necessarily vanishing.

The non-BPS D-branes with vanishing self-overlaps are given by  
the formula similar to  \eqn{desired brane 1}, 
but including the contribution from {\em the $a=2$ twisted sector\/};
\begin{eqnarray}
\dket{B} &=& \sqrt{2} \cP_4 \left[\dket{B}_0^{(a=0)} + \dket{B}_0^{(a=2)}\right]
\nn
& = &  \sqrt{2} \cP_2 \left[\dket{B}_0^{(\sNS, \, a=0)} + \dket{B}_0^{(\sR, \, a=2)}\right] 
\equiv
 \sqrt{2} \cP_2 \dket{B}^{\msc{(opp. BPS)}}_0,
\label{desired brane 2}
\end{eqnarray}
where 
\begin{equation}
\cP_4 \equiv \frac{1}{4}\sum_{n\in \bz_4}\, \sigma^n, \hspace{1cm} 
\cP_2 \equiv \frac{1}{2}\left( 1+ \sigma\right),
\end{equation}
and $\dket{B}_0^{(a=0)}$ is a BPS boundary state in the unorbifolded theory as before.
On the other hand, $\dket{B}_0^{(a=2)}$ is a suitably defined boundary state lying in the $a=2$ sector,
which contains the right-moving Ramond ground states with the opposite chirality as addressed above. 
(Obviously, we have no solutions of the boundary states in the $a=\pm 1$ sectors.) 
We return to this point shortly, but here just note 
$\sigma^2$ acts as $(-1)^{F_R-1}$ on the $a=2$ sector, rather than $(-1)^{F_R}$,
which is read off from the expression of $\f_{(a,b)}(\tau)$ in \eqn{fab2}.
The $a=2$ NSNS-sector is thus projected out by the $\cP_4$-action, while 
the $a=0$ RR-sector drops off.  We are then left with the 
$\cP_2$-projection of the `opposite BPS' boundary state,
which accounts for the second line of \eqn{desired brane 2}.
In other words, if we consider the type IIA (IIB) vacuum of 
this asymmetric orbifold, $\dket{B}^{\msc{(opp. BPS)}}_0$ 
is regarded as describing 
a BPS brane  in the type IIB (IIA) strings on $\br^{5,1}\times T^4[D_4]$.
One could schematically understand these aspects as 
\begin{align}
& [\mbox{IIA (IIB) vacuum on } T^4[D_4] / \sigma ] ~ \mbox{with} ~ \sigma^2 = (-1)^{F_R}
\nn
& \hspace{1cm}
\cong
[\mbox{IIB (IIA) vacuum on } T^4[D_4] / \sigma ] ~ \mbox{with} ~ \sigma^2 = {\bf 1}.
\label{IIA-IIB eq}
\end{align}

In fact, in the second case (ii), we can resolve the orbifold group as\footnote
   {If further  incorporating a shift operator into the orbifold action, {\em i.e.} considering the orbifolding by 
   $\sigma \otimes \cT_{2\pi R} $ 
as in \cite{SSW}, we do not have such a resolution.} 
\begin{equation}
\bz_4 ~ \mbox{generated by } \{ \sigma_{\msc{case (ii)}}\}  ~ \cong ~ \bz_2 \times \bz_2 ~
\mbox{generated by } \{ \sigma_{\msc{case (i)}}, ~ (-1)^{F_R} \},
\end{equation}
and by the relation suggested in \cite{GSen,Sen}, 
\begin{align}
[\mbox{IIA (IIB) vacuum}]/(-1)^{F_R} \cong [\mbox{IIB (IIA) vacuum}],
\end{align}
we obtain the above equivalence \eqn{IIA-IIB eq}.
Given this equivalence, a way to construct $\dket{B}_0^{(a=2)}$ is tracing back 
the relation in (\ref{desired brane 2}), as mentioned above.
The observation here is used to reduce the number of the cases to be analyzed 
in the following sections.

~


\subsection{Asymmetric Orbifold \ 
$\left[ T^4[D_2\oplus D_2] \times S^1_R\right] /\left[(\-_L)^{\otimes 2} \otimes (\-_R)^{\otimes 4} \right]$}
\label{subsec:torus2}

The point of the construction in the previous subsection is rather general
as described in the introduction, 
and various generalizations would be possible. 
As an example where the open-string boundary condition is more relevant, 
we next focus on a case of the 5-dim. torus $T^5$ along the $X^{5, \ldots, 9}$-directions.
To be more specific, we begin with the following compactification: 
\begin{itemize}
\item $X^{6,7,8,9}$-directions 

We consider
\begin{equation}
T^4[D_2 \oplus D_2] \equiv T^2[D_2] \times S^1[A_1] \times S^1[A_1],
\end{equation}
where $S^1[A_1]$ denotes the circle with the self-dual radius.

\item $X^5$-direction 

We just consider $S^1_R$, that is,  the circle compactification  
with an arbitrary radius $R$.

\end{itemize}
Then, we consider the orbifolding by 
\begin{equation}
\sigma := \left. (\-_L)^{\otimes 2}\right|_{5,6} \otimes \left. (\-_R)^{\otimes 4}\right|_{5,7,8,9},
\label{sigma 2}
\end{equation}
where $\left. (\-_L)^{\otimes 2}\right|_{5,6} $, for instance,  
means the chiral reflection acting along the left-movers of $X^{5, 6}$-directions.
Based on the twists of this type and related ones, 
non-SUSY vacua with vanishing cosmological constant have been investigated 
in \cite{SWada}.


The total modular invariant is given in the form, 
\begin{align}
Z(\tau,\bar{\tau}) & =  Z_{\msc{bosonic}}^{5d}(\tau,\bar{\tau}) \, \frac{1}{4} \sum_{a,b \in \bz_4}\, 
Z^{T^4\times S^1}_{(a,b)}(\tau,\bar{\tau})
 g_{(a,b)}(\tau) \overline{f_{(a,b)}(\tau)},
\label{total Z torus 2}
\end{align}
with
\begin{align}
Z^{T^4 \times S^1}_{(a,b)}(\tau,\bar{\tau})
& := 
\left\{
\begin{array}{ll}
 Z^{T^4 [D_2 \oplus D_2]}(\tau,\bar{\tau}) Z^{S^1_R}(\tau,\bar{\tau}) & ~~ (a,b \in 2\bz),
\\
\ep_{(a,b)}^{[2]} \chi^{D_2}_{(a,b)}(\tau)
\left| \chi^{A_1}_{(a,b)}(\tau) \right|^2 \left| \tchi^{A_1}_{(a,b)}(\tau) \right|^4 
\overline{\left(\tchi^{A_1}_{(a,b)}(\tau)\right)^2}
 & ~~ (a \in 2\bz+1, ~ \mbox{or} ~ b \in 2\bz+1).
\end{array}
\right.
\label{Z 2-1}
\end{align}
Here $Z^{5d}_{\msc{bosonic}}$ denotes the contribution from the bosonic  part of $\br^{4,1}$. 
In the second line, we have combined 
$ \big{|}\tchi^{A_1}_{(a,b)}(\tau)\big{|}^2$  from 
 the $X^5$-direction, $ \ep_{(a,b)}^{[-1]} \tchi^{A_1}_{(a,b)} \overline{\chi^{A_1}_{(a,b)}} $
 from the $X^6$-direction, and 
$\ep_{(a,b)}^{[3]} \chi^{D_2}_{(a,b)}\chi^{A_1}_{(a,b)}
\overline{\big(\tchi^{A_1}_{(a,b)}\big)^3}$ from the $X^{7,8,9}$-directions.
The character functions $\chi^{A_1}_{(a,b)}$, $\tchi^{A_1}_{(a,b)}$, $\chi^{D_2}_{(a,b)}$ and 
the free fermion chiral blocks $f_{(a,b)}$, $g_{(a,b)}$ are summarized in Appendix A. 
As already mentioned, the modular covariance of $Z^{T^4 \times S^1}_{(a,b)}$ is assured 
due to the phase factor $\ep^{[*]}_{(a,b)}$ \eqn{ep r}.


Again we have various possibilities of the action of $\sigma^2$ on the R-sector;
(i) $\sigma^2 ={\bf 1}$, (ii) $\sigma^2 = (-1)^{F_R}$, (iii) $\sigma^2 = (-1)^{F_L}$, 
(iv) $\sigma^2 = (-1)^{F_L+F_R}$. 
The modular invariant 
\eqn{total Z torus 2} 
describes the first case (i). 
The modular invariants for the remaining cases are easy to construct. Namely, 
we only have to replace the chiral blocks $f_{(a,b)}$,
$g_{(a,b)}$ in \eqn{Z 2-1} with the ones given in \eqn{fab2}, \eqn{gab2} suitably. 
However, as mentioned at the last part in the previous subsection, 
the cases (ii), (iii) reduces to the first case (i) as in \eqn{IIA-IIB eq},
and the case (iv) corresponds to a non-SUSY vacuum, which is beyond the scope of this work.

Therefore, it is enough to focus on the simplest case (i). 
We can pick up any BPS boundary states $\dket{B}_0$ in the unorbifolded theory 
on $T^4\times S^1$,  
and  define the non-BPS brane $\dket{B}$ by the orbifold projection 
in the same way as \eqn{desired brane 1}.
It is not difficult to show  that $\dket{B}$ has the vanishing self-overlap 
as long as $\dket{B}_0$ satisfies the general gluing condition (for the $T^4 \times S^1$-directions)
given by
\begin{align}
&\left[\al^5_{L , n} \pm \al^5_{R, -n} \right]\dket{B}_0 =0, 
&\left[\psi^5_{L, r} \pm i \psi^5_{R, -r} \right]\dket{B}_0 =0,
\nn
& \left[\al^6_{L,n} \pm \al^6_{R, -n} \right]\dket{B}_0 =0, 
&\left[\psi^6_{L, r} \pm i \psi^6_{R, -r} \right]\dket{B}_0 =0,
\nn
& \left[\al^i_{L, n} + M^i_{~j} \al^j_{R, -n} \right]\dket{B}_0 =0, 
& \left[\psi^i_{L, r} + i M^i_{~j} \psi^j_{R, -r} \right]\dket{B}_0 =0, 
\nn
& & \hspace{2cm}
(i,j = 7,8,9),
\label{glue cond torus 2}
\end{align}
where $\al^i_{L,n}$, $\al^i_{R,n}$ and $\psi^i_{L,r}$, $\psi^i_{R,r}$ denote the bosonic and fermionic 
oscillators (including the bosonic zero modes), and $M^i_{~j}$ is an arbitrary $SO(3)$-matrix.%
\footnote
   {Since $T^4[D_2 \oplus D_2] = T^4[(A_1)^4 ]$ holds, 
the $SO(3) (\subset SO(4))$-rotated gluing condition is  well-defined, even though $SO(3)$ 
is not a part of symmetry 
on this string vacuum.
}
To show this fact, it is useful to note the relation, 
\begin{equation}
\sigma \dket{B}_0 \equiv \left. (\-_L)^{\otimes 2} \right|_{5,6} \otimes \left. 
(\-_R)^{\otimes 4} \right|_{5,7,8,9} \dket{B}_0 
= \left. (\-_R)^{\otimes 4}\right|_{6,7,8,9} \dket{B}_0,
\end{equation}
for any $\dket{B}_0$ satisfying \eqn{glue cond torus 2}.
We thus obtain 
\begin{equation}
{}_0\dbra{B} \sigma e^{-\pi s H^{(c)}} \dket{B}_0 
\propto  f_{(0,1)}(is) = 0,
\label{Zcyl sigma torus 2}
\end{equation}
similarly to \eqn{Zcyl sigma torus 1}, leading to the vanishing cylinder amplitude, 
$
\dbra{B} e^{-\pi s H^{(c)}} \dket{B} =0.
$


~

We add a comment:

In the model of \eqn{subsec:torus1} 
the vanishing self-overlaps have been achieved for arbitrary  BPS boundary states  $\dket{B}_0$ 
in the unorbifolded theory.
On the other hand, in the current case,   $\dket{B}_0$ defined by \eqn{glue cond torus 2} 
is restricted to 
$(M^{i}_{~j}) \in SO(3)$  rather than $(M^{i}_{~j}) \in SO(4)$.
If adopting a different orbifold action instead of \eqn{sigma 2}, 
say, $ \sigma \equiv \left. (\-_L)^{\otimes 2}\right|_{5,7} \otimes \left. (\-_R)^{\otimes 4}\right|_{5,6,8,9}$,
we still obtain the same modular invariant, yielding the equivalent spectrum of closed string states. 
Moreover, it is obvious to have an essentially equivalent spectrum of non-BPS branes with 
the vanishing self-overlaps, in which 
$(M^{i}_{~j}) \in SO(3)$ appearing in \eqn{glue cond torus 2} should act on the $X_R^{6,8,9}$-directions 
this time. 
This fact is not surprising, of course. 
However, in the next section, we will see the examples in which the spectra of the non-BPS branes 
with vanishing self-overlaps would notably
depend on  the choice of orbifolding, while the modular invariant partition functions remain unchanged.

~


\section{Chiral SUSY Vacua as Asymmetric Orbifolds of $T^2 \times K3$}
\label{chiralsusyT2K3}

In this section, we  shall study less supersymmetric cases with $\cM = K3$ in the background 
\eqn{background 0},
\begin{equation}
\br^{3,1}\times T^2 [D_2]\times \cM.
\label{background 0'}
\end{equation}
The strategy to construct the non-BPS D-branes with vanishing self-overlaps 
is the same as in the previous section. 
The discussion is however a little more
involved. We thus first summarize the relevant asymmetric orbifolds in the bulk
in subsection \ref{orbifoldT2K3}. We then concentrate on the examples of 
the Gepner construction  in subsection \ref{Gepner}.
For these subsections, we follow \cite{KawaiS2} where the modular invariant 
partition functions of related asymmetric orbifolds (`mirrorfolds') are constructed.
(See also \cite{ES-G2orb}.)
Based on these set-ups, we construct the non-BPS D-branes in subsection \ref{N=4 bstate},
which is the main part of this section \ref{chiralsusyT2K3}.

\subsection{Asymmetric Orbifolds of $T^2 \times K3$ with Chiral SUSY}
\label{orbifoldT2K3}

To begin with, we
assume that the $\cM$-sector is described by a general $\cN=(4,4)$ SCFT with 
$\hat{c} \left( \equiv \frac{c}{3}\right)=2$, not reducing to the toroidal models.  
We denote the relevant  $\cN=4$ SCA 
\cite{Ademollo:1976wv}
by $L_n$ (Virasoro), $J^\alpha_n$ ($\widehat{SU}(2)_1$) with $\alpha=1,2,3$, 
$G^a_r$ with $a=0,1,2,3$.
Recall that the total R-symmetry is given by 
$SO(4) \cong SU(2)_c \times SU(2)_f $, in which the inner symmetry  
(`color $SU(2)$') is generated by the affine currents $J^\alpha$, whereas
the global $SU(2)$-symmetry  (`flavor $SU(2)$') is an outer one.
$G^0$ is a singlet of $SU(2)_{\msc{diag}} \subset SU(2)_c \times SU(2)_f$, while 
$G^1, G^2, G^3 $ compose a triplet of $SU(2)_{\msc{diag}}$.

We also assume that $\cN=2$ SCA is embedded into the $\cN=4$ one in the standard fashion by identifying
the $\cN=2$ $U(1)_R$-current as
\begin{equation}
J^{\cN=2} = 2 J^3,
\end{equation}
and 
\begin{equation}
G^{\pm} = \frac{1}{2} \left(G^0 \pm i G^3 \right).
\end{equation}
The generators of integral spectral flows $U_{\pm 1}$ are identified with the remaining $SU(2)$-currents  
\begin{equation}
U_{\pm 1} = J^{\pm} \equiv J^1 \pm i J^2,
\end{equation} 
and the half-spectral flows $U_{\pm 1/2}$ define the Ramond sector.


Let us now consider the asymmetric orbifolding of \eqn{background 0'} by 
$\sigma \equiv \sigma_{\cM} \otimes (\-_R)^{\otimes 2}$, 
where $(\-_R)^{\otimes 2}$ denotes the chiral reflection along the $T^2[D_2]$-direction. 
%
%
We first note the action of $(\-_R)^{\otimes 2}$ on the world-sheet fermions, 
which we assign to $\psi_R^4$, $\psi_R^5$. 
We bosonize them as\footnote
{If we had adopted the bosonization for the combinations, {\em e.g.} $ \psi_R^2 + i \psi_R^4$
and $\psi_R^3 + i \psi_R^5$, 
the L.H.S of \eqn{chiral ref square}
would have been the identity.
However, we shall not consider this possibility here in order to respect the super-Poincare symmetry in 
$\br^{3,1}$, and \eqn{bosonization T2} is the unique choice.
We also simply assume that $({\bf -1}_R)^{\otimes 2}$ is involutive on the bosonic coordinates 
$X^4_R$, $X^5_R$
(on the untwisted Hilbert space)
in this paper, even though we have more general possibilities if utilizing the fermionization of them 
as discussed in \cite{SWada}.
}
\begin{equation}
\psi_R^4 + i \psi_R^5 = \sqrt{2} e^{i H^{T^2}_R},
\label{bosonization T2}
\end{equation}
and $(\-_R)^{\otimes 2}$ should act as the shift $H^{T^2}_R \, \rightarrow \,  H^{T^2}_R + \pi$.
It fixes the action of $(\-_R)^{\otimes 2}$ on the R-sector and we find 
\begin{equation}
\left[(\-_R)^{\otimes 2}\right]^2 = (-1)^{F_R}.
\label{chiral ref square}
\end{equation}

We next consider 
the $\cM$-sector. 
We would like to suitably choose the orbifold twisting  $\sigma_{\cM}$ 
so as to obtain a 4-dim. vacuum with $\cN=(0,1)$-chiral SUSY unbroken.
Obviously $\sigma_{\cM}$ should be an automorphism of both left 
and right-moving $\cN=4$ SCAs.
Furthermore, since  working on superstring vacua in the NSR-formalism, $\sigma_{\cM}$ should 
satisfy the following conditions: 
\begin{description}
\item[(i)] 
$\sigma_{\cM}$ preserves $T$ (energy-momentum tensor) and $G^0$, 
which is necessary for the BRST-invariance.

\item[(ii)]
$\sigma_{\cM}$ keeps the Ramond sector intact so as to be compatible with $U_{\pm 1/2}$. 
This means that the automorphism $\sigma_{\cM}$ 
has to satisfy 
$
\sigma_{\cM} J^3 \sigma_{\cM}^{-1} = J^3,
$
or 
$
\sigma_{\cM} J^3 \sigma_{\cM}^{-1} = - J^3.
$
\end{description}
The same conditions are required for the right-mover.


Let us now introduce the automorphisms 
$\sigma_L^{(\alpha)}$ ($\alpha=1,2,3$)
of the left-moving $\cN=4$ SCA. They are defined by 
\begin{align}
& \sigma_L^{(\alpha)} T(z) \sigma_L^{(\alpha) -1} = T(z),  
\hspace{1cm} \sigma_L^{(\alpha)} G^0(z) \sigma_L^{(\alpha) -1} = G^0(z),
\nn
& \sigma_L^{(\alpha)} J^\alpha(z) \sigma_L^{(\alpha) -1} = J^\alpha(z), \hspace{1cm} 
\sigma_L^{(\alpha)} G^\alpha(z) \sigma_L^{(\alpha) -1} = G^\alpha(z),
\nn
& \sigma_L^{(\alpha)} J^\beta(z) \sigma_L^{(\alpha) -1} = - J^\beta(z), ~~ (\beta\neq \alpha), 
\hspace{1cm} \sigma_L^{(\alpha)} G^\beta(z) \sigma_L^{(\alpha) -1} = - G^\beta(z), ~~ (\beta\neq \alpha),
\label{def sigma i}
\end{align}
and we  assume that they are involutive on the whole Hilbert space;
$
\left(\sigma_L^{(\alpha)}\right)^2 = {\bf 1}_L 
$
$(\any \alpha)$.
We also set 
$
\hsigma_L^{(\alpha)} := e^{\frac{i\pi}{2} F_L} \sigma_L^{(\alpha)}
$
for convenience. $\hsigma_L^{(\alpha)}$ obviously acts on the $\cN=4$ SCA in the same way 
as \eqn{def sigma i}, but it is no longer involutive;
$
\left(\hsigma_L^{(\alpha)}\right)^2 = (-1)^{F_L}.
$

To complete the definition 
of $\sigma_L^{(\alpha)}$ (and $\hsigma_L^{(\alpha)}$),  
we still have to specify their actions on the Ramond ground states, in other words, 
on the half-spectral flow operators $U_{\pm 1/2}$.
Recalling the simple relation $J^{\pm} \equiv J^1\pm i J^2 = U_{\pm 1} = \left(U_{\pm 1/2}\right)^2$, 
we can naturally define 
\begin{equation}
\sigma_L^{(1)} \, U_{\pm 1/2} \,  \sigma_L^{(1) -1} = U_{\mp 1/2}, \hspace{1cm} 
\sigma_L^{(2)} \, U_{\pm 1/2} \,  \sigma_L^{(2) -1} = \pm i U_{\mp 1/2},
\label{sigma 1 2 Ramond}
\end{equation} 
which are surely consistent with 
$
\left(\sigma_L^{(\alpha)}\right)^2 = {\bf 1}_L.
$
We next consider the composition $\sigma^{(1)}_L \sigma^{(2)}_L$. 
It obviously acts on each $\cN=4$ chiral current in the same way as $\sigma^{(3)}_L$. 
However,  since $\sigma^{(1)}_L$ and $\sigma^{(2)}_L$ are anti-commutative on the R-sector 
due to \eqn{sigma 1 2 Ramond}, 
we find $(\sigma^{(1)}_L \sigma^{(2)}_L)^2 = (-1)^{F_L}$.
Thus, we should identify 
\begin{equation}
\sigma^{(1)}_L \sigma^{(2)}_L = (-1)^{F_L} \sigma^{(2)}_L \sigma^{(1)}_L 
= \hsigma^{(3)}_L \equiv e^{\frac{i\pi}{2} F_L} \sigma_L^{(3)},
\label{sigma rel}
\end{equation}
and $\hsigma^{(3)}_L $ acts on the half-spectral flows $U_{\pm 1/2}$ as 
\begin{equation}
\hsigma_L^{(3)} \, U_{\pm 1/2} \,  \hsigma_L^{(3) -1} =  \pm i U_{\pm 1/2}.
\label{hsigma 3 Ramond}
\end{equation}
$\sigma_R^{(\alpha)}$ ($\alpha=1,2,3$)  for the right-mover are defined in the same way.


Now, let us specify the possible orbifold actions 
$\sigma \equiv \sigma_{\cM} \otimes (\-_R)^{\otimes 2}$.
We again have four possibilities 
(i) $\sigma^2 = {\bf 1}$, (ii) $\sigma^2 = (-1)^{F_L}$, 
(iii) $\sigma^2 = (-1)^{F_R}$, (iv) $\sigma^2 = (-1)^{F_L+F_R}$, as in the previous section.
However, all the space-time SUSY are broken in  the fourth case, and the second and third cases 
reduce to the first case 
by the chirality flip; $\mbox{IIA} \, \longleftrightarrow\, \mbox{IIB}$
as mentioned in 
subsection
\ref{subsec:torus1}.
It is thus enough to consider the first case such that $\sigma $ is involutive, that is,
the cases with  $\sigma \equiv \sigma^{(\alpha)}_L \otimes \hsigma^{(\beta)}_R \otimes (\-_R)^{\otimes 2}$.
We shall especially focus on the following 
three cases;
(1) $\sigma_{\cM} \equiv \sigma^{(3)}_L \otimes \hsigma^{(1)}_R $, 
(2) $\sigma_{\cM} \equiv \sigma^{(3)}_L \otimes \hsigma^{(3)}_R $,
(3) $\sigma_{\cM} \equiv \sigma^{(1)}_L \otimes \hsigma^{(1)}_R $. 
Of course, we have to examine whether they are actually compatible with the modular invariance. 
In the next subsection, we explicitly  confirm 
in the case of the Gepner construction
that the asymmetric orbifolding by $\sigma \equiv \sigma_{\cM} \otimes (\-_R)^{\otimes 2}$
constructed this way yields superstring vacua with  modular invariance.  
In all the three cases, the space-time SUSY from the left mover is broken
to achieve the 4-dim. $\cN =(0,1)$ chiral SUSY.

~


\subsection{Concrete Examples : Gepner Construction }
\label{Gepner}


Let us consider the generic Gepner construction \cite{Gepner} for 
$K3$, that is, 
the superconformal system defined by
\begin{eqnarray}
 && \left\lb M_{k_1}\otimes \cdots  \otimes
M_{k_r}\right\rb\left|_{\bsz_N\msc{-orbifold}}
\right. ~, ~~~ \sum_{i=1}^r \frac{k_i}{k_i+2}=2~,
\label{Gepner K3}
\end{eqnarray}
where $M_k$ denotes the $\cN=2$ minimal model of level $k$
($\hat{c}\equiv \frac{c}{3}= \frac{k}{k+2}$), and we 
set
\begin{eqnarray}
 N := \mbox{L.C.M.} \{k_i+2~;~i=1,\ldots,r\}.
 \label{def N}
\end{eqnarray}
We start with the diagonal modular invariant for simplicity.
We have to make the $\bz_N$-orbifolding that renders the total $U(1)_R$-charge 
(in the NS-sector) integral,
\begin{eqnarray}
&& Q(I) :=  \sum_{i=1}^r \frac{m_i}{k_i+2} \in \bz,
\label{U(1) charge constraint} 
\end{eqnarray}
where 
$I :=  \{ (\ell_1,m_1), \ldots, (\ell_r,m_r)\}$
denotes the collective label of the primary state in $M_{k_1}\otimes \cdots \otimes M_{k_r}$
 ($0\leq \ell_i \leq k_i$, 
$m_i \in \bz_{2(k_i+2)}$, $\ell_i+m_i \in 2\bz$),
and the twisted sectors of orbifolding are identified with the `spectral flow orbits' by the actions $\cU^n$ 
($n \in \bz_N$) 
with
\begin{eqnarray}
 \cU \, :\,  I\equiv \{(\ell_1,m_1),\cdots, (\ell_r,m_r)\}~ \longmapsto ~
 \cU (I)\equiv \{(\ell_1,m_1-2),\cdots, (\ell_r,m_r-2)\}.
\end{eqnarray}
See \cite{EOTY} for more detail.


The relevant  Hilbert space for the $K3$ sector (before imposing the GSO projection)
is schematically expressed as  
\begin{eqnarray}
 && \cH^{(s,\tilde{s})}_{\msc{Gepner}} = \bigoplus_{n\in \bsz_N} 
\bigoplus_{\stackrel{\scriptstyle I,\tI}{Q(I)\in \bz,~ Q(\tI) \in \bz}} \,\left\lb 
\delta_{I,\tI}
\, \cH^{(s)}_{\cU^n(I),L} \otimes 
\cH^{(\tilde{s})}_{\tI,R} \right\rb , ~~~ (s, \tilde{s} =\NS, \R),
\label{H Gepner}
\end{eqnarray}
where the Ramond Hilbert space $\cH^{(R)}_{I, *}$ is uniquely determined by the half-spectral flow 
in the standard manner.\footnote
  {Notice however that the label $I$ in $\cH^{(R)}_{I, *}$ indicates the quantum numbers 
  in the {\em NS-sector.}  }
Note that the left-right symmetric primary states lie in the  
$n=0$ sector, but we also have many asymmetric primary states generated
by the spectral flows. 
As already mentioned, 
the $\cN=2$ superconformal symmetry with $\hat{c}=2$ is enhanced to the $\cN=4$ 
by adding the spectral flow operators, which are identified with 
the $\widehat{SU}(2)_1$ currents $J^{\pm} \equiv J^1\pm i J^2$ in the $\cN=4$
SCA \cite{EOTY}. Accordingly, the chiral parts of 
$\cH^{(s,\tilde{s})}_{\msc{Gepner}}$  are decomposed into irreducible
representations of $\cN=4$ SCA at level 1, that are classified as follows 
\cite{ET}:
\begin{itemize}
 \item {{\bf massive representations:} $\cC^{(\sNS)}_h$, $\cC^{(\sR)}_h$}

These are non-degenerate representations whose vacua have   
conformal weights $h$. 
The vacuum of $\cC^{(\sNS)}_h$
belongs to the spin $0$ representation of the $\widehat{SU}(2)_1$-symmetry. 
The four-fold degenerate vacua of $\cC^{(\sR)}_h$ 
generate the representation $2[\mbox{spin 0}] \oplus [\mbox{spin 1/2}]$.
Unitarity requires $h\geq 0$ for $\cC^{(\sNS)}_h$ and $h\geq
\frac{1}{4}$ for $\cC^{(\sR)}_h$.
The 1/2-spectral flow connects $\cC^{(\sNS)}_h$ with $\cC^{(\sR)}_{h+\frac{1}{4}}$.
 \item {{\bf massless representations:} $\cD^{(\sNS)}_{\ell}$, 
$\cD^{(\sR)}_{\ell}$  ($\ell=0,1/2$)}

These are degenerate representations whose vacua have conformal weights 
$h=\ell$ for the NS representations $\cD^{(\sNS)}_{\ell}$, and 
$h=\frac{1}{4}$ for the Ramond representations $\cD^{(\sR)}_{\ell}$;
they belong to the spin $\ell$ representation of $\widehat{SU}(2)_1$.
To be more specific, $\cD^{(\sNS)}_0$
(`graviton rep.' or `identity rep.') corresponds to the unique vacuum
with $h=0$, $J^3_0=0$, while $\cD^{(\sNS)}_{1/2}$ (`massless matter rep.') is
generated over doubly degenerated vacua with $h=1/2$, $J^3_0=\pm 1/2$.
The Ramond sector $\cD^{(\sR)}_{\frac{1}{2}-\ell}$ is connected with $\cD^{(\sNS)}_{\ell}$
by the 1/2-spectral flow.

\end{itemize}
The relevant character formulas are summarized in Appendix A.



Now, let us construct the asymmetric orbifolds by the involution $\sigma$.
Since a detailed account of closely related asymmetric orbifolds has been
given in \cite{KawaiS2}, based on \cite{ES-G2orb}, 
we here briefly describe the relevant construction.

Since the most non-trivial part $\sigma_{\cM}$ has the form 
$\sigma^{(\alpha)}_L \otimes \hsigma^{(\beta)}_R$ ($\alpha,\beta = 1 ~  \mbox{or} ~ 3$),
we should specify how the $\cN=4$ involutions  $\sigma^{(1)}_L$, $\sigma^{(3)}_L$ 
act on the primary states in the Gepner construction. 
First, 
we can naturally identify $\sigma^{(1)}_L$ with the $\cN=2$ involution,
\begin{equation}
\sigma^{(1)}_L : = \prod_{i=1}^r \sigma^{\cN=2, (i)}_L,
\label{sigma1L} 
\end{equation}
where the $\cN=2$ involution $\sigma^{\cN=2, (i)}_{L}$ acts as 
\begin{align}
& T^{(i)} \, \rightarrow \, T^{(i)}, ~~~ J^{(i)} \, \rightarrow \, - J^{(i)}, 
~~~ G^{\pm \, (i)} \, \rightarrow \, G^{\mp\, (i)},
\label{N=2 invoution}
\end{align}
in each minimal factor $M_{k_i}$.

On the other hand, $\sigma^{(3)}_{L}$ acts on the $\cN=4$ SCA as the automorphism \eqn{def sigma i}. 
We still have to define how it acts on the $\cN=4$ primary states $\ket{v}_L$.
A simple choice would be given as 
\begin{equation}
\sigma^{(3)}_{L} \ket{v}_L := \left\{
\begin{array}{ll}
\sigma^{(1)}_L \ket{v}_L~, &
(2J^3_{L,0}\ket{v}_L=0)~, \\
J^+_{L,0} \, \sigma^{(1)}_L \ket{v}_L~, &
(2J^3_{L,0}\ket{v}_L=\ket{v}_L)~, \\
- J^-_{L,0} \, \sigma^{(1)}_L 
\ket{v}_L~, &
(2J^3_{L,0}\ket{v}_L=-\ket{v}_L)~,
\end{array}\right.
\label{sigma 3 primary}
\end{equation}
where $J^{\pm}_L \equiv J^1_L \pm i J^2_L$ are the $SU(2)$
currents in the $\cN=4$ SCA,
which turns out to be compatible with the modular invariance.

By these definitions and the fact that $\sigma^{(1)}_{L\,(R)}$ and $\sigma^{(3)}_{L\,(R)}$
induce the equal twisted characters of $\cN=4$ SCA (see Appendix B), we find that the torus partition function  
does not depend on $\alpha$, $\beta$ in 
$\sigma_{\cM} \equiv \sigma^{(\alpha)}_L \otimes \hsigma^{(\beta)}_R$.
The total modular invariant is now  written as  
\begin{align}
Z(\tau,\bar{\tau}) & : = Z^{4d}_{\msc{bosonic}}(\tau,\bar{\tau}) 
\, \frac{1}{4} \sum_{a,b \in \bz_4}\, Z_{(a,b)}(\tau,\bar{\tau}) .
\label{Ztotal Gep}
\end{align}
As before, $Z^{4d}_{\msc{bosonic}}$ denotes the contribution from the bosonic  part of $\br^{3,1}$, 
which is 
related with neither the $\sigma$-twisting nor the GSO-projection.
Those for the various $\sigma$-twisted sectors  $Z_{(a,b)}$ ($a,b\in \bz_4$), 
which are crucial in our arguments,  are described in the following way: 
\begin{itemize}
\item {\bf even sectors with $a,b \in 2\bz$} 
\begin{align}
Z_{(a,b)}(\tau,\bar{\tau}) & : = \frac{1}{4} \sum_{s,\tilde{s}} \, \sum_{\cI,\tcI}\, N_{\cI,\tcI}\, F^{(s)}_{\cI}(\tau) \overline{F^{(\tilde{s})}_{\tcI}(\tau)} 
\cdot Z^{T^2[D_2]}(\tau,\bar{\tau}) \cdot 
\left(\frac{\th_{[s]}}{\eta} \right)^2 \overline{ \left(\frac{\th_{[\tilde{s}]}}{\eta} \right)^2},
\label{Gep Zab even}
\end{align}
where $F^{(s)}_{\cI}(\tau)$, $F^{(\tilde{s})}_{\tcI}(\tau)$ denote the chiral building blocks 
with the chiral spin structures $s, \tilde{s} = \NS,\,\tNS, \, \R, \, \tR, $
in the Gepner model for $\cM$, which are labeled by the spectral flow orbits 
$\cI$, $\tcI$.  
For instance,  $F^{(\sNS)}_{\cI}(\tau)$ is explicitly written as 
$$
F^{(\sNS)}_{\cI}(\tau) = \sum_{\{(\ell_i, m_i)\} \in \cI} \, \prod_{i=1}^r\, \ch{(\sNS)}{\ell_i,m_i}(\tau),
$$
with 
$$
\cI \equiv  \{(\ell_1, m_1 -2n), \ldots , (\ell_r, m_r-2n)\}_{n \in \bz_N},
$$
and the $\cN=2$ minimal character $ \ch{(s)}{\ell_i,m_i}(\tau)$ \cite{Dobrev,RY1}.
The chiral blocks for other spin structures are determined by the $1/2$-spectral flows and by incorporating 
 the suitable sign factors to impose the GSO condition. 
(See \cite{EOTY} for more detail.)
We also adopted the concise notation $\th_{[s]}(\tau) :=  \th_3(\tau), ~ \th_4(\tau), 
~ \th_2(\tau), ~ i\th_1(\tau)(\equiv 0)$ for $s=  \NS,\,\tNS, \, \R, \, \tR$
respectively. 
The modular invariant coefficients $N_{\cI, \tcI}$ are straightforwardly determined 
due to the Gepner construction, which are independent of $a,b$,
and the overall factor $1/4$ originates from the chiral GSO projection. 


\item {\bf odd sectors with $a\in 2\bz+1$ or $b\in 2\bz+1$} 
\begin{align}
Z_{(a,b)}(\tau,\bar{\tau}) & : = 
Z^{\cM}_{(a,b)}(\tau,\bar{\tau}) \cdot Z^{T^2[D_2]}_{(a,b)}(\tau,\bar{\tau}) \cdot Z^f_{(a,b)}(\tau,\bar{\tau})
\nn
 & \equiv \sum_{\bl, \tbl}\,
 N^{(a,b)}_{\bl,\tbl}\, \chi_{\bl, (a,b)}(\tau) \overline{\chi_{\tbl, (a,b)}(\tau)}
 \cdot 
\ep^{[2]}_{(a,b)} 
\chi^{D_2}_{(a,b)}(\tau) \overline{\left( \tchi^{A_1}_{(a,b)}(\tau)\right)^2}
\nn
& \hspace{1cm}
\times 
\ep^{[2]}_{(a,b)} \chi^{D_2, [-]}_{(a,b)}(\tau) 
\frac{1}{2} \overline{\left[\left( \tchi^{A_1}_{(a,b)}(\tau)\right)^2 
- \left( \tchi^{A_1}_{(a,b)}(\tau)\right)^2 \right]},
\label{Gep Zab odd}
\end{align}
where we set
\begin{align}
& \chi_{\bl, (a,b)}(\tau) :=  \prod_{i} \, \chi^{k_i}_{\ell_i, [a,b]}(\tau),
\hspace{1cm}  \bl \equiv (\ell_1, \ldots, \ell_r ) ,
\end{align}
and $\chi^{k}_{\ell, [a,b]}(\tau)$ denotes the twisted $\cN=2$ character \eqn{twisted minimal}.
Recall that 
$
\ep^{[r]}_{(a,b)}\equiv e^{\frac{i\pi}{8}r (-1)^a ab}
$
and the definitions of the functions 
$\chi^{D_2}_{(a,b)}(\tau)$, $\chi^{D_2, [-]}_{(a,b)}(\tau)$ and $\tchi^{A_1}_{(a,b)}(\tau)$ 
are summarized in \eqn{Dr ab}, \eqn{Dr - ab} and \eqn{tA1 ab}. 
The 4-dim. $\cN =(0,1)$ chiral SUSY is confirmed from \eqn{Gep Zab odd}.


The coefficients $N^{(a,b)}_{\bl,\tbl}$ in the odd sectors are slightly non-trivial. 
We can determine them in a 
way parallel to that presented in \cite{KawaiS2,ES-G2orb}.    
We here briefly describe the results, which depend on the spectrum of the level $k_i$ 
in \eqn{Gepner K3} as follows: 
\begin{description}
\item[(i) At least one of $k_i$'s is odd] 

~

In this case, the modular invariant coefficients are very simple, 
\begin{eqnarray}
 N^{(a,b)}_{\bl,\tbl}= \prod_{i=1}^r\delta_{\ell_i,
\tilde{\ell}_i}.
\label{N a b 1}
\end{eqnarray}


\item[(ii) All $k_i$'s are even]

~

In this case,  $N$  in \eqn{def N} is  even,  and we set
\begin{eqnarray}
 && S_1 := \Big\{i\in \{1,\ldots, r\} ~;~ \frac{N}{k_i+2}\in 2\bz+1\Big\}, \nn
 && S_2 := \Big\{i\in \{1,\ldots, r\} ~;~ \frac{N}{k_i+2}\in 2\bz\Big\}.
\end{eqnarray}
Then, the relevant coefficients are given by
\begin{equation}
N^{(a,b)}_{\bl,\tbl} : = \left\{
\begin{array}{ll}
\dsp  \prod_{i\in S_2}
\delta_{\ell_i,\tell_i} 
\left( \prod_{i\in S_1}\delta_{\ell_i,\tell_i}
+\prod_{i\in S_1} \delta_{\ell_i,k_i-\tell_i}\right) 
& ~~ (a\in 2\bz, ~ b\in 2\bz+1),
\\
\dsp 
\left(1+(-1)^{\sum_{i\in S_1}\ell_i}\right)
\prod_{i=1}^r\delta_{\ell_i,\tell_i}
& ~~ (a \in 2\bz+1).
\end{array}
\right.
\label{N a b 2}
\end{equation}
\end{description}

\end{itemize}

One can directly confirm that the $Z_{(a,b)}(\tau, \bar{\tau})$ in the odd sectors \eqn{Gep Zab odd} 
show the suitable 
modular covariance by using the modular transformation formulas given in \eqn{modular twisted minimal}.
Note that $\sigma^{(\al)}_{L \, (R)}$-insertion only provides non-vanishing contributions 
to the trace over the sectors  with 
$\{ (\ell_1, 0), \, \ldots, \, (\ell_r, 0)\}$ in the spectral flow orbit (of NS-sector).
The difference of the two cases \eqn{N a b 1} and \eqn{N a b 2} originates from this fact. 

~


We make a few comments: 
\begin{itemize}
\item As already mentioned, we are considering the orbifolding by 
$\sigma = \sigma^{(\alpha)}_L \otimes \hsigma^{(\beta)}_R \otimes (\-_R)^{\otimes 2}$
for various $\alpha$, $\beta$, and 
obtain the equivalent spectra of closed string states in all these models. 
However, this fact does not necessarily imply that they are equivalent string vacua. Indeed, 
it turns out that they have quite different 
D-branes,
as we elucidate in subsection \ref{N=4 bstate}.

\item  
One finds that the contributions from the  $(\R, *)$ or $(*,\R)$-sectors do not appear 
in the building block \eqn{Gep Zab odd} with $a \in 2\bz$ and 
$b\in 2\bz+1$. 
This fact is actually expected so as to achieve the modular invariance.
It is not difficult to confirm that this is indeed the case in almost all the Gepner models for $K3$
due to the basic properties of the twisted characters. (See Appendix B.)
The exception is only the $(4)^3$ type, in which there would exist a non-vanishing 
$(\R, \NS)$ contribution\footnote
   {The $(\R,\R)$ and $(\NS,\R)$-contributions trivially vanish due to the fermionic zero-modes 
   in the $\br^{3,1} \times T^2[D_2]$-directions. 
}  that could spoil the modular invariance. 
However, by  suitably fixing the sign ambiguity of $\sigma^{(\alpha)}_L$ on the Ramond vacua with $Q=0$, 
one can avoid this possibility still in the $(4)^3$-model. 

\end{itemize}

~


\subsection{General Construction of Boundary States with Vanishing Self-overlaps}
\label{N=4 bstate}

Let us present our main studies. Namely, we  investigate how we achieve the vanishing self-overlaps 
in the current models of asymmetric orbifolds, 
\begin{equation}
\br^{3,1} \times \left. \left[T^2[D_2] \times \cM \right] \right|_{\sigma-\msc{orbifold}}.
\label{def bg}
\end{equation}

We begin with specifying the boundary conditions in the (unorbifolded) $\cM$-sector 
that characterize  general BPS D-branes.\footnote
   {In this paper, we shall not work with explicit forms of the boundary states in Gepner model
   for  $\cM$, which 
 should be constructed as tensor products of those for the $\cN=2$ minimal models. 
See \cite{Recknagel:1997sb} and also {\em e.g.}  \cite{Brunner:1999jq}, \cite{GSatoh}  for detail. 
}
Naively, any boundary conditions preserving  the $\cN=4$ superconformal symmetry 
with an arbitrary twisting by automorphism
may be allowed,
which are schematically expressed as in \cite{OOY} by
\begin{equation}
\left[ \cA^I_{r} + g\cdot \widetilde{\cA}^I_{-r} \right] \dket{B} =0 . 
\label{N=4 bc 0}
\end{equation}
Here, $\cA^I_r$, $\widetilde{\cA}^I_{r}$ are the chiral currents and 
$g$ denotes any (inner or outer) automorphism of the $\cN=4$ SCA.
However, since we are working on the physical boundary states in the RNS superstrings, 
we still have to impose the following conditions: 
\begin{description}
\item[(i)] 
$\dket{B}$  preserves $G^0$-symmetry without any twisting, which is necessary for the BRST-invariance.

\item[(ii)]
$\dket{B}$ contains the correct components of the RR-sector 
compatible with the above definition of $U_{\pm 1/2}$. This means that the automorphism $g$ 
in \eqn{N=4 bc 0} has to satisfy 
$
g \cdot J^3 = J^3,
$
or 
$
g \cdot J^3 = -J^3.
$
\end{description}
Thus, at least generically,  the allowed twisting $g$ by the $\cN=4$ automorphism is restricted and we 
eventually obtain the following two types of gluing conditions: 
\begin{description}
 \item[A-type : ]  
\begin{eqnarray}
&& 
\left[ L_n-\tL_{-n}\right] \dket{\th}_A=0, 
~~~ \left[ J^3_n-\tJ^3_{-n}\right] \dket{\th}_A=0, \nn
&& 
\left[ G_r^0-i\tG_{-r}^0\right] \dket{\th}_A=0, ~~~
\left[ G_r^3+i\tG_{-r}^3\right] \dket{\th}_A=0,  \nn
&& 
\left[G_r^\alpha - i \widehat{R}(\th)^\alpha_{~\beta}\tG_{-r}^\beta\right] \dket{\th}_{A}=0,~~~
 \left[ J_n^\alpha +  \widehat{R}(\th)^\alpha_{~\beta} \tJ_{-n}^\beta\right] \dket{\th}_{A}=0,
 ~~~ (\alpha,\beta=1,2). \qquad
\label{N=4 gluing A}
\end{eqnarray}
 \item[B-type : ]  
\begin{eqnarray}
&& \left[ L_n-\tL_{-n}\right] \dket{\th}_B=0, ~~~ \left[ J^3_n+\tJ^3_{-n}\right] \dket{\th}_B=0, \nn
&& \left[ G_r^\alpha-i\tG_{-r}^\alpha\right] \dket{\th}_B=0, ~~(\alpha=0,3)~, ~\nn
&& 
\left[ G_r^\alpha-i R(\th)^\alpha_{~\beta}\tG_{-r}^\beta\right] \dket{\th}_B=0,~~~
 \left[ J_n^\alpha + R(\th)^\alpha_{~\beta} \tJ_{-n}^\beta\right] \dket{\th}_B=0,~~~ (\alpha,\beta=1,2). \qquad
\label{N=4 gluing B}
\end{eqnarray}

\end{description}
In the above, 
$R(\th)$ denotes the $SO(2)$-rotation with the angle parameter $\th$, and
$\widehat{R}(\th) \equiv R(\th)\sigma_3 \in O(2)$.
The relevant Ishibashi states \cite{Ishibashi}
are characterized by the $\cN=4$ irrep. classified in the  subsection \ref{Gepner}  
as well as the gluing conditions  
given above, and should satisfy {\em e.g.}
\begin{align}
&  {}_A \dbra{\cD^{(\sNS)}_{\ell} ; \th} 
e^{-\pi s H^{(c)}} \dket{ \cD^{(\sNS)}_{\ell}  ; \th}_A = \ch{(\sNS)}{0}(\ell; is),
\nn
& {}_A \dbra{\cC^{(\sNS)}_h ; \th} e^{-\pi s H^{(c)}} \dket{\cC^{(\sNS)}_h; \th}_A 
= \ch{(\sNS)}{}(h; is) \equiv e^{-2\pi \left( h-\frac{1}{8}\right)}
\frac{\th_3(is)^2}{\eta(is)^3} ,
\end{align}
where 
$\ch{(\sNS)}{0}(\ell;is)$, 
$\ch{(\sNS)}{}(h; is)$ denote the $\cN=4$ massless and massive  characters
summarized in \eqn{N=4 massive}, \eqn{N=4 massless matter} and \eqn{N=4 grav}. 
To be more precise, 
since the Gepner points are rational, 
it turns out that only the discrete values of the angle parameter 
$\th =  \frac{2\pi   r}{N}$ ($r\in \bz_N$) are allowed. 
In fact, let us recall the schematic decomposition of an $\cN=4$ irrep. by the integral spectral flows as 
$$
[\mbox{irrep.}]^{(\sNS), \, \cN=4} = \bigoplus_{n \in \bz_N}\, U_n \, [\mbox{irrep.}]^{(\sNS), \, \cN=2},
$$
where $N$ is defined by  \eqn{def N}, and 
we also express the $\cN=2$ Ishibashi state of the A-type 
as $\dket{[\mbox{irrep.}]^{(\sNS)}}_A^{\cN=2}$ (defined by the gluing conditions given 
in the first and second lines in
\eqn{N=4 gluing A}).
Then, the $\cN=4$ Ishibashi states of A-type with the twist angle $\th= \frac{2\pi r}{N} $ are written  as  
\begin{align}
\dket{[\mbox{irrep.}]^{(\sNS)};\th=  \frac{2\pi r}{N} }_A 
= \sum_{n\in \bz_N}\, (-1)^n e^{2\pi i \frac{r}{N} n} \, U_{n} 
\otimes \widetilde{U}_{n} \dket{[\mbox{irrep.}]^{(\sNS)
 }}^{\cN=2}_A.
\label{N=4 Ishibashi}
\end{align}
This shows why $\th$ is restricted to discrete values $\th= \frac{2\pi r}{N} $. 
The B-type Ishibashi states are similarly constructed. 

The Ishibashi states in the RR-sector are obtained by the half-spectral flow from the NSNS ones, 
\begin{align}
& \dket{[\mbox{irrep.}]^{(\sR)};\frac{2\pi r}{N} }_A 
= U_{1/2} \otimes \widetilde{U}_{1/2} \dket{[\mbox{irrep.}]^{(\sNS)}; \frac{2\pi r}{N}}_A,
\nn
&
\dket{[\mbox{irrep.}]^{(\sR)};\frac{2\pi r}{N} }_B 
= U_{1/2} \otimes \widetilde{U}_{-1/2} \dket{[\mbox{irrep.}]^{(\sNS)}; \frac{2\pi r}{N} }_B.
\label{def dket R}
\end{align}
We note the correspondence of the representations, 
\begin{align}
& U_{\pm 1/2} ~ : ~ \cD^{(\sNS)}_{\ell} ~ \longrightarrow ~  \cD^{(\sR)}_{1/2-\ell}, ~~~ (\ell=0, 1/2),
\nn
& U_{\pm 1/2} ~ : ~ \cC^{(\sNS)}_{h} ~ \longrightarrow ~  \cC^{(\sR)}_{h+\frac{1}{8}}.
\end{align}
The R-massive rep. $\cC^{(\sR)}_{h}$ is generated by doubly degenerated vacua with conformal weight $h$
belonging to an $SU(2)$-doublet, as opposed to the NS-one $\cC^{(\sNS)}_{h} $.


As in the previous analyses on the toroidal models, 
generic D-branes in our asymmetric orbifold \eqn{def bg} 
are expressed by the boundary states in the form 
of the orbifold projection with $\sigma^2=1$,
\begin{equation}
\dket{B} = \sqrt{2} \cP \dket{B}_0 
\equiv  \sqrt{2} \cP \left[\dket{B}^{(\sNS)}_0 + \dket{B}^{(\sR)}_0\right],
\label{bstate general}
\end{equation}
where $\dket{B}_0$ is a (GSO-projected) boundary state describing a D-brane in the unorbifolded theory
and $\cP \equiv \frac{1+\sigma}{2}$. 
We assume that $\dket{B}_0$ describes a half-BPS brane with the Dirichlet conditions 
for all the transverse coordinates along $\br^{3,1} \times T^2[D_2]$, just for convenience.
Namely, $\dket{B}_0$ is expanded by the Ishibashi states given above for the $\cM$-sector and 
the self-overlap is schematically written as 
\begin{align}
{}_0\dbra{B} 
e^{-\pi s H^{(c)}} \dket{B}_0 
 & 
=  \sum_{i} \, \al_i \, 
\frac{1}{\eta^4} \left[ \left(\frac{\th_3}{\eta}\right)^2 \ch{(\sNS)}{*}(r^{(\sNS)}_i;is) 
-\left(\frac{\th_4}{\eta}\right)^2 \ch{(\stNS)}{*}(r^{(\sNS)}_i;is) 
\right.
\nn
& \hspace{2cm}
-  \left.  
\left(\frac{\th_2}{\eta}\right)^2 \ch{(\sR)}{*}(r^{(\sR)}_i ; is)
\right] \equiv 0,
\label{cyl 0}
\end{align}
where $r_i^{(\sNS)}$ and $r_i^{(\sR)}$ are unitary  irrep.'s 
of $\cN=4$ SCA related with each other 
by $U_{\pm 1/2}$ and 
$\al_i$ are some non-trivial coefficients
that we are not interested in here.
The R.H.S of \eqn{cyl 0} indeed vanishes due to the BPS-property of $\dket{B}_0$. 
One can easily confirm that the each term associated to the irrep. $r_i^{(*)}$ 
actually vanishes.

Therefore, to achieve the vanishing cylinder amplitudes in the asymmetric orbifolds \eqn{def bg}, 
it is enough to examine  whether or not the amplitude
$
{}_0\dbra{B} \sigma e^{-\pi s H^{(c)}} \dket{B}_0 
$
vanishes. 
From now on, we examine this problem in each case of 
(1) $\sigma_{\cM} \equiv \sigma^{(3)}_L \otimes \hsigma^{(1)}_R $, 
(2) $\sigma_{\cM} \equiv \sigma^{(3)}_L \otimes \hsigma^{(3)}_R $,
(3) $\sigma_{\cM} \equiv \sigma^{(1)}_L \otimes \hsigma^{(1)}_R $, 
as addressed before. 
We set 
$\th_r \equiv \frac{2\pi r}{N}$, ($r\in \bz_N$) in the following.

~

\noindent
 (1) $\sigma_{\cM} = \sigma^{(3)}_L \otimes \hsigma^{(1)}_R$ :

We first pick up the $\cM$-sector.
Because of the gluing conditions \eqn{N=4 gluing A}, \eqn{N=4 gluing B}, we obtain the equality 
\begin{align}
\sigma_{\cM} \dket{*; \th_r }_{A (B)} \equiv \sigma^{(3)}_L \otimes \hsigma^{(1)}_R \dket{*; \th_r}_{A (B)} 
= \hsigma^{(1)}_R  \sigma^{(3)}_R \dket{*; \th_r}_{A (B)}
= \sigma^{(2)}_R  \dket{*; \th_r}_{A (B)}.
\label{id bstate sigma 1}
\end{align}
It is worthwhile to emphasize that this relation does not depend on the angle parameter $\th_r$ at all. 
Thus, the amplitude from each component of Ishibashi state 
is eventually evaluated by the $\sigma^{(2)}_R$-twist 
irrespective of $\th_r$,
yielding   the $\cN=4$ twisted  character, 
\begin{equation}
\chi_{[0,1]}(h; is) \equiv \frac{2e^{- 2\pi s \left(h-\frac{1}{8}\right)}}{\th_2(is)},
\label{chi01} 
\end{equation}
or trivially vanishing one.
We summarize necessary formulas for the $\cN=4$ twisted characters in Appendix B.
In this way, we obtain for the NSNS-sector, 
\begin{align}
\hspace{-1cm}
 {}_{A (B)} \dbra{\cD^{(\sNS)}_{0} ; \th_r} 
\sigma_{\cM} 
e^{-\pi s H^{(c)}} \dket{ \cD^{(\sNS)}_{0}  ; \th_r}_{A (B)} 
& = 
{}_{A (B)} \dbra{\cD^{(\sNS)}_{0} ; \th_r} 
(-1)^{f_L} \sigma_{\cM} 
e^{-\pi s H^{(c)}} \dket{ \cD^{(\sNS)}_{0}  ; \th_r}_{A (B)} 
\nn
&= 
\chi_{[0,1]}(h=0; is),
\nn
\hspace{-1cm}
 {}_{A (B)} \dbra{\cD^{(\sNS)}_{1/2} ; \th_r} 
\sigma_{\cM} 
e^{-\pi s H^{(c)}} \dket{ \cD^{(\sNS)}_{1/2}  ; \th_r}_{A (B)} 
& = 
{}_{A (B)} \dbra{\cD^{(\sNS)}_{1/2} ; \th_r} 
(-1)^{f_L} \sigma_{\cM} 
e^{-\pi s H^{(c)}} \dket{ \cD^{(\sNS)}_{1/2}  ; \th_r}_{A (B)} 
\nn
& =0,
\nn
\hspace{-1cm}
{}_{A (B)} \dbra{\cC^{(\sNS)}_h ; \th_r} 
\sigma_{\cM} 
e^{-\pi s H^{(c)}} 
\dket{\cC^{(\sNS)}_h; \th_r}_{A (B)} 
&
=
{}_{A (B)} \dbra{\cC^{(\sNS)}_h ; \th_r} 
(-1)^{f_L}
\sigma_{\cM} 
e^{-\pi s H^{(c)}} 
\dket{\cC^{(\sNS)}_h; \th_r}_{A (B)} 
\nn
&  = \chi_{[0,1]} (h; is) ,
\end{align}
where $(-1)^{f_L}$ denotes the twisting for the GSO projection.
The fact that $(-1)^{f_L} \sigma_{\cM}$-insertion leads to the equal amplitude 
is obvious from the boundary conditions for the fermionic currents $G^a(z)$, ($a=0,1,2,3$).

We also recall that $\sigma$ includes $(\-_R)^{\otimes 2}$, which just makes  
the free fermion contribution from the (transverse part of) $\br^{3,1} \times T^2[D_2]$-sector
proportional to $\dsp \frac{\th_3}{\eta}\frac{\th_4}{\eta}$
for the NSNS-sector, while $(-1)^{f_L} (\-_R)^{\otimes 2}$ gives 
the term
$\dsp \propto \frac{\th_4}{\eta}\frac{\th_3}{\eta}$.
On the other hand, the contributions from the RR-sector trivially vanish due to free fermion zero-modes 
along either of the $\br^{3,1}$ or $T^2[D_2]$-directions.

Combining all the contributions and taking account of the GSO-projection, we finally obtain 
\begin{align}
{}_0\dbra{B} \sigma 
e^{-\pi s H^{(c)}} \dket{B}_0 
 & 
=  \sum_{i} \, \al'_i \, 
\frac{1}{\eta^4} \left[ \frac{\th_3}{\eta}\frac{\th_4}{\eta} - \frac{\th_4}{\eta}\frac{\th_3}{\eta}\right]
 \chi_{[0,1]}(h_i; is)\equiv 0.
\label{cyl 1}
\end{align}
In this expression\footnote
  {We note that the coefficients $\al_i'$ are not necessarily equal to those appearing in 
\eqn{cyl 0}, since they would depend on the phases arising from the $\sigma_{\cM}$-actions 
on the $\cN=4$ primary states.    
} the summation is taken over all the spin 0 irrep.'s,  
that is, $\cC^{(\sNS)}_h$ or $\cD^{(\sNS)}_0$, 
and we assign $h_i=0$ for the case of $\cD^{(\sNS)}_0$.
In this way, we have shown  that any boundary states 
\eqn{bstate general} associated to $\dket{B}_0$ satisfying  the gluing conditions \eqn{N=4 gluing A} or 
\eqn{N=4 gluing B} {\em with an arbitrary value of parameter} $\th_r= \frac{2\pi r}{N}$  $(r\in \bz_N)$  
provide the vanishing self-overlaps, 
\begin{align}
\dbra{B} 
e^{-\pi s H^{(c)}} \dket{B} = 0.
\end{align}

As in the toroidal case in section \ref{toroidal models}, 
the couplings of $\dket{B}$ and the closed string states are multiplied 
by the overall factor in $\dket{B}$. The D-brane tension and the RR charge
are hence $\sqrt{2}$ times those in the unorbifolded theory.
The open string excitations in the unorbifolded theory
remain in the self-overlap of $\dket{B}_0$, which are tachyon-free.

~

\noindent
(2)
$\sigma_{\cM} = \sigma^{(3)}_L \otimes \hsigma^{(3)}_R$ :

In the second case, 
\eqn{id bstate sigma 1} should be replaced with 
\begin{align}
\sigma_{\cM} \dket{*; \th_r}_{A (B)} \equiv 
\sigma^{(3)}_L \otimes \hsigma^{(3)}_R \dket{*; \th_r}_{A (B)} 
= \hsigma^{(3)}_R  \sigma^{(3)}_R \dket{*; \th_r}_{A (B)}
= e^{\frac{i \pi}{2} F_R}  \dket{*; \th_r}_{A (B)}.
\label{id bstate sigma 2}
\end{align}
Thus, the net effect of the twist is just a phase factor for the RR-component of boundary state. 
Incorporating also the $\br^{3,1} \times T^2$-sector, 
the RR-component of the overlap again drops off due to the fermionic zero-modes, and 
we obtain the following 
amplitude instead of \eqn{cyl 1}, 
\begin{align}
{}_0\dbra{B} \sigma
e^{-\pi s H^{(c)}} \dket{B}_0 
 & 
=  \sum_{i} \, \al_i \, 
\frac{1}{\eta^4} \left[ \frac{\th_3}{\eta} \frac{\th_4}{\eta}\ch{(\sNS)}{*}(r^{(\sNS)}_i;is) 
-\frac{\th_4}{\eta} \frac{\th_3}{\eta} \ch{(\stNS)}{*}(r^{(\sNS)}_i;is) 
\right].
\label{cyl 2}
\end{align}
At least for generic Gepner models, the R.H.S of \eqn{cyl 2} does not vanish 
for any value of the moduli parameter $\theta_r$.
In fact, R.H.S of \eqn{cyl 2} does not depend on $\th_r$, and
$$
\ch{(\sNS)}{*}(r^{(\sNS)}_i;\tau) \neq \ch{(\stNS)}{*}(r^{(\sNS)}_i;\tau),
$$
for a generic rep. $r_i$.
Rephrasing more physically, the D-brane tension has been modified by the $\sigma$-insertion, 
while the RR-charge remains the same as in case (1).
This causes the mismatch of amplitudes for the graviton and RR-particle exchanges.
In this way, we conclude that all of  the D-branes in the second case have non-vanishing self-overlaps, 
as one expects from the general features of non-BPS D-branes.


~

\noindent
(3)
$\sigma_{\cM} = \sigma^{(1)}_L \otimes \hsigma^{(1)}_R$ :

The third case is the most subtle one. 
When translating the $\sigma^{(1)}_L$-insertion into that of the right-mover 
similarly to \eqn{id bstate sigma 1}, \eqn{id bstate sigma 2}, 
we have to take account of the $R(\th)$ ($\widehat{R}(\th)$) rotation appearing 
in the gluing conditions \eqn{N=4 gluing B} (\eqn{N=4 gluing A}). 
For instance, for 
the B-type gluing condition, we obtain 
\begin{align}
\sigma_{\cM} \dket{*; \th_r}_{B} \equiv 
\sigma^{(1)}_L \otimes \hsigma^{(1)}_R \dket{*; \th_r}_{B} 
= \hsigma^{(1)}_R   \sigma^{(1), [\th_r]}_R \dket{*; \th_r}_{B}, 
\label{id bstate sigma 3}
\end{align}
instead of \eqn{id bstate sigma 1}, \eqn{id bstate sigma 2}, where 
$ \sigma^{(1), [\th_r]}_R $ denotes the automorphism acting on the $\cN=4$ SCA rotated by $R(\th_r)$
in the same way as $\sigma^{(1)}_R$.\footnote
   {Since the $R(\th_r)$-rotation is an outer-automorphism, it seems difficult 
to write $\sigma^{(1), [\th_r]}_R $ down explicitly. }
Obviously the relation \eqn{id bstate sigma 3} yields the self-overlap that depends on the parameter $\th_r$,
as opposed to the first and second cases. 
The resultant amplitude does not vanish generically. 
However, for the special value $\th_r= \pm \frac{\pi}{2}$, we find 
\begin{equation}
\hsigma^{(1)}_R  \sigma^{(1), [\pm \frac{\pi}{2}]}_R =  \hsigma^{(1)}_R   \sigma^{(2)}_R 
= (-1)^{F_R} \sigma^{(3)}_R,
\end{equation}
yielding the cancellation as given in \eqn{cyl 1}.
The A-type gluing condition is likewise treated.

In this way, we conclude that the D-branes in the third case 
have the vanishing self-overlaps only for the gluing conditions with 
$\th_r \equiv \frac{2\pi r}{N} = \pm \frac{\pi}{2}$,
which is possible when $N \in 4\bz_{>0}$.

~


\noindent
{\bf Absence or presence of tachyonic instabilities}

Here we would like to further 
discuss whether the non-BPS branes considered above could include 
the tachyonic instabilities. 
Since it is obvious that no closed string tachyons appear in the relevant boundary states, 
we should examine the open string excitations in the orbifolded sector. 
Indeed, it is easy to estimate  the lightest excitation in the open string channel.
By detailed case studies, it would be possible to write down the formulas 
of the general spectra, which are however beyond the scope of this paper.

Let us  first note common features  in the orbifolded sector
for the above three cases;
(i) the RR contribution to the self-overlap vanishes due to the fermionic zero-modes, implying the lack of GSO-projection for the open string Hilbert space,
(ii) the twist by $(\-_R)^{\otimes 2}$ along the $T^2[D_2]$-direction  
adds the conformal weight
$
\frac{1}{4}
$
to the open string vacua.

Now, the estimations for the above three cases are summarized as follows; 
\begin{description}
\item[case (1) : ]

In this case we have the bose-fermi cancellation in the open string spectrum as noted above.
Thus, it is enough to consider the NS-sector.

Recall that $\sigma_{\cM}$ acts on the $\cN=4$ primary states as the  product of 
the $\cN=2$ involutions for each minimal sector $M_{k_i}$,
which gives rise to the energy shifts bounded from below by
$ \frac{\hat{c}_i}{8} \equiv \frac{k_i}{8(k_i+2)}$ in the open string spectrum. 
(See the formula of conformal weight \eqn{h t}.)
Eventually we find that the minimum value of conformal weight 
for the open string excitations 
should satisfy the inequality;
\begin{equation}
h^{(\msc{min})} \geq \frac{1}{4} + \sum_{i=1}^r \, \frac{\hat{c}_i}{8} = \frac{1}{2},
\label{hmin 1}
\end{equation}
and the inequality can be saturated only when all the $k_i$'s are even. 
Therefore, the lightest open string excitation could be massless when all $k_i$'s are even, 
and always massive if at least some $k_i$'s are odd. 
In this way, we conclude that no tachyonic instability emerges in the open string spectrum. 


\item[case (2) : ]

$\sigma_{\cM}$ again acts on the $\cN=4$ primary states
in the same way, whereas it effectively makes the $\cN=4$ SCA invariant, 
after taking account of the identity
\eqn{id bstate sigma 2}.
Thus, the twisted $\cN=4$ character 
$\chi_{[0,1]}(*; is) \propto \frac{\th_3\th_4}{\eta^3} (is) $ for the case (1) 
has to be replaced with the untwisted one $\propto \frac{\th_3^2}{\eta^3} (is) $
for the NS-sector.
Making the modular transformation,  the net effect  just amounts to  the shift by 
$ - \frac{1}{8}$ 
to the R.H.S of
\eqn{hmin 1}.
We thus obtain the inequality
\begin{equation}
h^{(\msc{min})} \geq \frac{1}{4} + \left\{ \sum_{i=1}^r \, \frac{\hat{c}_i}{8} - \frac{1}{8}\right\} 
= \frac{3}{8},
\label{hmin 2}
\end{equation}
and open string tachyons would appear. 
This result is  expected since the open string spectrum is non-supersymmetric in this case.

\item[case (3) : ]

Again, $\sigma_{\cM}$ acts on the $\cN=4$ primary states 
as the above two cases. 
On the other hand, by utilizing  \eqn{id bstate sigma 3}, we find that the net effect on the (right-moving) $\cN=4$ SCA 
by the $\sigma_{\cM}$-insertion amounts to  the $SO(2)$-rotation with the angle parameter $2\th_r$ on  the $J_R^1$, $J_R^2$ (and $G_R^1$, $G_R^2$) plane,
while leaving the other generators intact. 
Then, the twisted $\cN=4$ character 
$ \propto \frac{\th_3\th_4}{\eta^3} (is) $ for the case (1) 
is replaced with
$
\propto \frac{\th_3(is, 0) \th_3(is , 2\th_r) }{\eta^3} ,
$
which induces the additional energy shift of the amount: $ -\frac{1}{8} + \frac{1}{8\pi^2}(2\th_r)^2$ 
to the R.H.S of \eqn{hmin 1}.
The relevant inequality now becomes 
\begin{equation}
h^{(\msc{min})} \geq \frac{1}{4} + \left\{ \sum_{i=1}^r \, \frac{\hat{c}_i}{8} 
- \frac{1}{8} + \frac{1}{8\pi^2}(2\th_r)^2 \right\} 
= \frac{3}{8} + \frac{\th_r^2}{2 \pi^2}.
\label{hmin 3}
\end{equation}
This implies that open string tachyons would generically emerge except for the special angle 
$ \th_r= \frac{\pi}{2}$
for $N \in 4\bz_{>0}$, which realizes the bose-fermi cancellation in the open string spectrum 
as mentioned above.

\end{description}



~

\subsection{Points of Toroidal Orbifolds}
\label{sec.ZNpt}

Our discussion so far is based mostly only on general properties of the $\cN=4$ SCFT for $\cM$.
Thus, we would expect 
that the spectrum of the non-BPS branes with the vanishing overlaps is unchanged 
over generic points of the moduli space of $K3$, 
as long as the asymmetric orbifolding by $\sigma \equiv   (\-_R)^{\otimes 2} \otimes  \sigma_{\cM} $ 
is well-defined.
The points in our argument were: 
\begin{itemize}
\item The global symmetry $SU(2)_{\msc{diag}}$ 
preserving $G^0$
is only identified with an {\em outer}-automorphisms of the $\cN=4$ SCA.

\item We need to pick up a particular $U(1)$-subalgebra of the $\cN=4$ SCA to define the Ramond sector
by the half-spectral flows,
which has been generated by $J^3$ in the above arguments.

\end{itemize}
Then, only the restricted $SO(2) ( \subset SU(2)_{\msc{diag}})$ twisting is 
allowed in the gluing conditions \eqn{N=4 gluing A}, \eqn{N=4 gluing B}, 
so as to preserve the Ramond sector Hilbert space.

On the other hand, there are special points with the `symmetry enhancement' in the moduli space, 
at which more general gluing conditions could be solved. 
For instance, it has been known \cite{EOTY} that 
the Gepner model $(2)^4$ (Kummer surface)
is equivalent with the  $\bz_2$-orbifold of  $T^4[D_4, B_{ij} \equiv 0]$, which is defined as  the 4-dim. torus 
associated to the root lattice of $D_4$ with the vanishing Kalb-Ramond field.\footnote{
To avoid a possible confusion,
we here emphasize that $T^4[D_4, B_{ij} \equiv 0]$ differs from the symmetry enhancement point of 
$\widehat{SO}(8)_1$, which is denoted as `$T^4[D_4]$', say, in \eqn{Z D4} 
in the present paper (and also \cite{SatohS}).
}
We can reinterpret this system in terms of free bosons and fermions,  
and thus, the $SU(2)_{\msc{diag}}$ is explicitly realized by these free fields. 
In this special case all the choices of orbifold twisting 
$\sigma_{\cM} = \sigma_L^{(\al)}\otimes \hsigma_R^{(\beta)}$ ($\al,\beta=1,2,3$)
lead to equivalent superstring vacua, as in the toroidal models studied in section \ref{toroidal models}. 
Especially we find the equivalent spectra of the non-BPS D-branes 
with the vanishing self-overlaps. 
Indeed, with the help of free field interpretation, one can straightforwardly solve the following
equations for the boundary states, 
\begin{eqnarray}
&& 
\hspace{-1cm}
\left[ L_n-\tL_{-n}\right] \dket{\th, \varphi}=0, 
~~~ \left[ G_r^0-i\tG_{-r}^0\right] \dket{\th, \varphi}=0, \nn
&& 
\hspace{-1cm}
\left[G_r^a - i R(\th, \varphi)^a_{~b}\tG_{-r}^b\right] \dket{\th, \varphi}=0,~~~
 \left[ J_n^a + R(\th, \varphi)^a_{~b} \tJ_{-n}^b\right] \dket{\th, \varphi}=0,~~~ (a,b=1, 2, 3),
\label{N=4 gluing special}
\end{eqnarray}
where $R(\th, \varphi)$ denotes an arbitrary $SO(3)$-rotations.

There also exist the $\bz_3$, $\bz_4$, $\bz_6$-orbifold points within the Gepner models 
for $K3$ as discussed in \cite{EOTY}.
However,  such an enhancement of symmetry does not happen for these points, 
and $SU(2)_{\msc{diag}}$ is still identified as  outer-automorphisms. 

~


\section{Discussion}

We have studied the type II string vacua with chiral space-time SUSY constructed 
as asymmetric orbifolds, focusing on the D-branes on these backgrounds. 
The simple but crucial idea in this paper is  
that all the D-branes are non-BPS in any chiral SUSY vacua.
As clarified in sections 2 and 3, one can straightforwardly construct  
the chiral SUSY vacua based on asymmetric orbifolds 
which accommodate rather generally the non-BPS D-branes with vanishing cylinder amplitudes.
This would be hardly realized in the geometrical compactifications of superstring theory.

We have especially investigated the asymmetric orbifolds of $T^2 \times \cM$, 
as well as simpler toroidal models,  
where $\cM= K3$ is described by 
a general $\cN=4$ SCFT with $c=6$ defined by the Gepner construction.
We have demonstrated in subsection \ref{N=4 bstate}
 that the spectra of such non-BPS D-branes with the bose-fermi cancellation depend notably 
 on the choice of orbifolding, 
even when the closed string spectra remain unchanged. 
This feature is in contrast to those of the toroidal asymmetric orbifolds presented 
in section \ref{toroidal models}.

In this respect we note that the most of the analyses on the boundary states 
given in  subsection \ref{N=4 bstate}  are based only 
on general properties  of the $\cN=4$ SCFT for $\cM$, as mentioned in the previous section.
Thus, the spectrum of the non-BPS D-branes with vanishing cylinder amplitudes 
would be unchanged over generic points of the moduli space of $K3$, as long as
the asymmetric twist is well-defined. 
The point in our discussion is summarized in subsection  \ref{sec.ZNpt}.
The exception would be the orbifold point 
with symmetry enhancement.



Based on the results in this paper, one may now discuss 
a possible application to the problem of cosmological constant. 
As mentioned in the introduction, the cosmological constant induced solely 
by the non-BPS D-branes would be exponentially suppressed for small string coupling.
Furthermore, in a given non-BPS D-brane background, the contributions to the 
closed-string vacuum amplitude 
would come only from the diagrams with the external legs sourced by that non-BPS D-brane.
The analysis of the loops thus would be much simpler than the case of the bulk SUSY-breaking
\cite{Kachru1,Kachru3,Aoki:2003sy,Abel:2017rch}, to control the almost vanishing cosmological
constant. It would also be challenging to substantiate the scenario \cite{Harvey}, which is 
based on the analysis on the heterotic dual side and mentioned in the introduction,
that the non-BPS D-branes condensate to produce the non-perturbative mismatch
of the spectrum.
This would also be an interesting  problem involving a non-supersymmetric duality.
We hope to return to these issues elsewhere.



\section*{Acknowledgments}
This work is supported in part by JSPS Grant-in-Aid for Scientific Research
24540248 and 17K05406, Japan-Hungary Research Cooperative Program and 
Japan-Russia Research Cooperative Program 
from Japan Society for the Promotion of Science (JSPS).

~


\appendix

\section*{Appendix A: ~ Summary of  Conventions}

\setcounter{equation}{0}
\def\theequation{A.\arabic{equation}}

\noindent
\underline{\bf Theta functions} 
%
 \begin{align}
 & \dsp \th_1(\tau,z):=i\sum_{n=-\infty}^{\infty}(-1)^n q^{(n-1/2)^2/2} y^{n-1/2}
  \equiv  2 \sin(\pi z)q^{1/8}\prod_{m=1}^{\infty}
    (1-q^m)(1-yq^m)(1-y^{-1}q^m), \nn [-10pt]
   & \\[-5pt]
 & \dsp \th_2(\tau,z):=\sum_{n=-\infty}^{\infty} q^{(n-1/2)^2/2} y^{n-1/2}
  \equiv 2 \cos(\pi z)q^{1/8}\prod_{m=1}^{\infty}
    (1-q^m)(1+yq^m)(1+y^{-1}q^m), \\
 & \dsp \th_3(\tau,z):=\sum_{n=-\infty}^{\infty} q^{n^2/2} y^{n}
  \equiv \prod_{m=1}^{\infty}
    (1-q^m)(1+yq^{m-1/2})(1+y^{-1}q^{m-1/2}),  
\\
 &  \dsp \th_4(\tau,z):=\sum_{n=-\infty}^{\infty}(-1)^n q^{n^2/2} y^{n}
  \equiv \prod_{m=1}^{\infty}
    (1-q^m)(1-yq^{m-1/2})(1-y^{-1}q^{m-1/2}) . 
\\
& \Th{m}{k}(\tau,z):=\sum_{n=-\infty}^{\infty}
 q^{k(n+\frac{m}{2k})^2}y^{k(n+\frac{m}{2k})} ,
\\
&
\eta(\tau) := q^{1/24}\prod_{n=1}^{\infty}(1-q^n).
 \end{align}
 Here, we have set $q:= e^{2\pi i \tau}$, $y:=e^{2\pi i z}$  
 ($\any \tau \in \bh^+$, $\any z \in \bc$),
 and used abbreviations, $\th_i (\tau) \equiv \th_i(\tau, 0)$
 ($\th_1(\tau)\equiv 0$), 
$\Th{m}{k}(\tau) \equiv \Th{m}{k}(\tau,0)$.

~


\noindent
\underline{\bf Bosonic building blocks}

Here we summarize the notation of the building blocks 
used in the main text according to \cite{SatohS}.
%
%
Associated to the basic representation of  $\widehat{(D_r)}_1$  ($r \in 2\bz_{>0}$), we set  
\begin{align}
 \chi^{D_r}_{(a,b)}(\tau) 
& :=
\left\{
\begin{array}{ll}
\frac{1}{2 \eta(\tau)^r}
\left\{\th_3(\tau)^r + e^{\frac{i\pi r}{4} a} \th_4(\tau)^r\right\},
& ~~ (a\in 2\bz, ~ b\in 2\bz+1), \\
\frac{1}{2 \eta(\tau)^r}
\left\{\th_3(\tau)^r + e^{\frac{i\pi r}{4} b} \th_2(\tau)^r\right\},
& ~~ (a\in 2\bz+1, ~ b\in 2\bz), \\
\frac{1}{2 \eta(\tau)^r}
\left\{\th_4(\tau)^r + e^{\frac{i\pi r}{4} (a+b-1)} \th_2(\tau)^r\right\},
& ~~ (a\in 2\bz+1, ~ b\in 2\bz+1). \\
\end{array}
\right.
\label{Dr ab}
\end{align}
We also define the following functions, 
\begin{align}
 \chi^{D_r, [-]}_{(a,b)}(\tau) 
& :=
\left\{
\begin{array}{ll}
\frac{1}{2 \eta(\tau)^r}
\left\{\th_3(\tau)^r - e^{\frac{i\pi r}{4} a} \th_4(\tau)^r\right\},
& ~~ (a\in 2\bz, ~ b\in 2\bz+1), \\
\frac{1}{2 \eta(\tau)^r}
\left\{\th_3(\tau)^r - e^{\frac{i\pi r}{4} b} \th_2(\tau)^r\right\},
& ~~ (a\in 2\bz+1, ~ b\in 2\bz), \\
\frac{1}{2 \eta(\tau)^r}
\left\{\th_4(\tau)^r - e^{\frac{i\pi r}{4} (a+b-1)} \th_2(\tau)^r\right\},
& ~~ (a\in 2\bz+1, ~ b\in 2\bz+1), \\
\end{array}
\right.
\label{Dr - ab}
\end{align}
which are associated to the vector representation of  $\widehat{(D_r)}_1$.


For $(\widehat{A_1})_1$,  we introduce 
\begin{align}
 \chi^{A_1}_{(a,b)}(\tau) 
&:= 
\left\{
\begin{array}{ll}
\dsp
\frac{1}{2}
\left\{
\chi^{A_1}_+(\tau) + e^{\frac{i\pi}{2} a} \chi^{A_1}_-(\tau)
\right\},
& ~~ (a\in 2\bz, ~ b\in 2\bz+1), \\
\dsp 
\frac{1}{\sqrt{2}}
\left\{
\chi^{A_1}_0 (\tau) + e^{\frac{i\pi}{2} b} \chi^{A_1}_1(\tau)
\right\},
& ~~ (a\in 2\bz+1, ~ b\in 2\bz), \\
\dsp
\frac{1}{\sqrt{2}}
\left\{
\chi^{A_1}_0 (\tau) + e^{\frac{i\pi}{2}(a+ b-1) } \chi^{A_1}_1(\tau)
\right\},
& ~~ (a\in 2\bz+1, ~ b\in 2\bz+1), \\
\end{array}
\right.
\label{A1 ab}
\end{align}
where we set 
\begin{equation}
\chi^{A_1}_{\pm}(\tau) := \chi^{A_1}_0(\tau) \pm  \chi^{A_1}_1(\tau),
\end{equation}
and the $(\widehat{A_1})_1$-characters are given as 
\begin{align}
& \chi^{A_1}_0(\tau) 
:= \frac{\th_3(2\tau)}{\eta(\tau)} \equiv \frac{\Th{0}{1}(\tau)}{\eta(\tau)},
~~~ (\mbox{basic rep.}),
\nn
& \chi^{A_1}_1(\tau) 
:= \frac{\th_2(2\tau)}{\eta(\tau)} \equiv \frac{\Th{1}{1}(\tau)}{\eta(\tau)},  
~~~ (\mbox{spin $1/2$ rep.}).
\label{A1 ch}
\end{align}
%
%
On the other hand, 
we define 
\begin{align}
 \tchi^{A_1}_{(a,b)}(\tau) 
&:=
\left\{
\begin{array}{ll}
\dsp 
\sqrt{\frac{\th_3(\tau)\th_4(\tau)}{\eta(\tau)^2}},
& ~~ (a\in 2\bz, ~ b\in 2\bz+1),\\
\dsp 
\sqrt{\frac{\th_3(\tau)\th_2(\tau)}{\eta(\tau)^2}},
& ~~ (a\in 2\bz+1, ~ b\in 2\bz),\\
\dsp 
\sqrt{\frac{\th_4(\tau)\th_2(\tau)}{\eta(\tau)^2}},
& ~~ (a\in 2\bz+1, ~ b\in 2\bz+1), \\
\end{array}
\right.
\label{tA1 ab}
\end{align}
which are interpretable as 
the $\widehat{(A_1)}_1$-characters twisted by the involution 
$
\rho^{(\al)}_{A_1} \equiv e^{- i\pi \frac{\ell}{2}} e^{i\pi J^{\al}_0},
$
($\al=1,2,3$)
for the spin $\ell/2$-integrable representation of $\widehat{(A_1)}_1$.


~

\noindent
\underline{\bf Fermionic building blocks}

To describe the supersymmetric chiral blocks for the free fermions, 
we introduce the notation
\begin{equation}
\cJ(\tau) := \frac{1}{2\eta(\tau)^4} \left\{\th_3(\tau)^4 - \th_4(\tau)^4 - \th_2(\tau)^4 \right\} \left(\equiv 0 \right),
\label{cJ}
\end{equation}
and associated to the reflection of four components $(\-_L)^{\otimes 4}$,
\begin{eqnarray}
 f_{(a,b)}(\tau)
&:=  &  q^{\frac{1}{4}a^2}e^{\frac{i\pi}{2}ab}
\,
\left(\frac{\th_1\left(\tau,\frac{a\tau+b}{2}\right)}{\eta(\tau)}\right)^2
\left(\frac{\th_1(\tau,0)}{\eta(\tau)}\right)^2
\nn
&\equiv &
\left\{
\begin{array}{ll}
\dsp
 e^{\frac{i\pi}{2}ab} \frac{1}{2 \eta(\tau)^4}
\left\{
\th_3(\tau)^2\th_4(\tau)^2
- \th_4(\tau)^2\th_3(\tau)^2
+0 \right\},
 &  ~~ (a\in 2\bz,~ b\in 2\bz+1),\\
\dsp
 e^{\frac{i\pi}{2}ab} \frac{1}{2 \eta(\tau)^4}
\left\{
\th_3(\tau)^2 \th_2(\tau)^2 +0 
- \th_2(\tau)^2\th_3(\tau)^2
\right\},
 &  ~~ (a\in 2\bz+1,~ b\in 2\bz),\\
\dsp
-  e^{\frac{i\pi}{2}ab} \frac{1}{2 \eta(\tau)^4}
\left\{ 0+
\th_2(\tau)^2 \th_4(\tau)^2
- \th_4(\tau)^2\th_2(\tau)^2
\right\},
 &  ~~ (a\in 2\bz+1,~ b\in 2\bz+1),\\
\dsp \cJ(\tau)
& ~~ (a\in 2\bz, ~ b\in 2\bz). 
\end{array}
\right.
\nn
&&
\label{fab}
\end{eqnarray}
In the second line, each term corresponds to the NS, $\tNS$, R sectors with  keeping this order.
These trivially vanish, as is consistent with the space-time SUSY. 
They satisfy the modular covariance of the form,
\begin{eqnarray}
 && f_{(a,b)}(\tau)|_S \equiv f_{(a,b)}\left(-\frac{1}{\tau}\right)
= f_{(b,-a)}(\tau), \nn
&& f_{(a,b)}(\tau)|_T \equiv f_{(a,b)}(\tau+1)
= - e^{-2\pi i \frac{1}{6}} f_{(a,a+b)}(\tau) .
\label{mc fab}
\end{eqnarray}


We next define the non-supersymmetric chiral block twisted  
by the two component reflection $(\-_L)^{\otimes 2}$, 
\begin{align}
 g_{(a,b)}(\tau) 
&
:= 
(-1)^{ab}
\ep^{[-2]}_{(a,b)} 
\left[ \tchi^{A_1}_{(a,b)}(\tau) \right]^2 \chi^{D_2, [-]}_{(a,b)}(\tau) 
\nn
&\equiv 
\left\{
\begin{array}{ll}
\dsp  e^{-\frac{i\pi}{4}ab}
 \frac{1}{2\eta(\tau)^4}
\left\{
\th_3(\tau)^3
\th_4(\tau)
- (-1)^{\frac{a}{2}} 
\th_4(\tau)^3
\th_3(\tau)
+0
\right\},
& ~~ (a\in 2\bz, ~ b\in 2\bz+1)\\
\dsp
 e^{\frac{i\pi}{4}ab}
 \frac{1}{2\eta(\tau)^4}
\left\{
\th_3(\tau)^3
\th_2(\tau)
+0 
- (-1)^{\frac{b}{2}} 
\th_2(\tau)^3
\th_3(\tau)
\right\},
& ~~ (a\in 2\bz+1, ~ b\in 2\bz)\\
\dsp
 - e^{\frac{i\pi}{4}ab}
 \frac{1}{2\eta(\tau)^4}
\left\{
0+
\th_4(\tau)^3
\th_2(\tau)
+i (-1)^{\frac{a+b}{2}} 
\th_2(\tau)^3
\th_4(\tau)
\right\},
& ~~ (a\in 2\bz+1, ~ b\in 2\bz+1) \\
\cJ(\tau)
 & ~~ (a \in 2\bz, ~ b\in 2\bz),
\end{array}
\right.
\label{gab}
\end{align}
and also for the twisting by $(-1)^{F_L}$,
\begin{align}
h_{(a,b)}(\tau) & :=  q^{\frac{a^2}{2}}e^{i\pi ab}
\left(\frac{\th_1\left(\tau, \frac{a\tau+b}{2}\right)}{\eta(\tau)}\right)^4
\nn
& \equiv
\left\{
\begin{array}{ll}
\dsp
\frac{1}{2\eta(\tau)^4} \left\{\th_3(\tau)^4 - \th_4(\tau)^4 + \th_2(\tau)^4 \right\}  \equiv 
\left(\frac{\th_2(\tau)}{\eta(\tau)}\right)^4 ,
& ~~ (a \in 2\bz, ~ b\in 2\bz+1 ),
\\
\dsp
\frac{1}{2\eta(\tau)^4} \left\{\th_3(\tau)^4 + \th_4(\tau)^4 - \th_2(\tau)^4 \right\}  \equiv 
\left(\frac{\th_4(\tau)}{\eta(\tau)}\right)^4 ,
& ~~ (a \in 2\bz+1, ~ b\in 2\bz ),
\\
\dsp
- \frac{1}{2\eta(\tau)^4} \left\{\th_3(\tau)^4 + \th_4(\tau)^4 + \th_2(\tau)^4 \right\}  \equiv 
- \left(\frac{\th_3(\tau)}{\eta(\tau)}\right)^4 ,
& ~~ (a , b \in 2\bz+1 ),
\\
 \cJ(\tau),  & ~~ (a, b \in 2\bz).
\end{array}
\right.
\label{hab}
\end{align}
Again they satisfy the modular covariance in the same sense as \eqn{mc fab}.


We also introduce slightly modified chiral blocks, 
\begin{align}
\f_{(a,b)}(\tau) & :=
\left\{
\begin{array}{ll}
f_{(a,b)}(\tau), & ~~ (a \in 2\bz+1 ~ \mbox{or} ~ b \in 2\bz+1),
\\
h_{(\frac{a}{2}, \frac{b}{2})}, & ~~ (a, b \in 2\bz), 
\end{array}
\right. 
\label{fab2}
\end{align}
\begin{align}
\g_{(a,b)}(\tau) & :=
\left\{
\begin{array}{ll}
g_{(a,b)}(\tau), & ~~ (a \in 2\bz+1 ~ \mbox{or} ~ b \in 2\bz+1),
\\
h_{(\frac{a}{2}, \frac{b}{2})}, & ~~ (a, b \in 2\bz).
\end{array}
\right. 
\label{gab2}
\end{align}
They correspond to the cases 
of $\left[(\-_L)^{\otimes 4} \right]^2 = (-1)^{F_L}$, and $\left[(\-_L)^{\otimes 2} \right]^2 = (-1)^{F_L}$, respectively, 
and behave modular covariantly as above. 

~

\noindent
\underline{\bf Characters for the $\cN=4$ SCA with $c=6$} 

The character formulas of the unitary  irrep.'s 
of the $\cN=4$ SCA with $c=6$ (level 1) are given in 
\cite{ET}, and we exhibit them here. We focus on the NS-sector: 
\begin{description}
 \item[massive representation $\cC^{(\sNS)}_h$] 
\begin{eqnarray}
  \ch{\cN=4,(\sNS)}{}(h;\tau,z) &=& q^{h-\frac{1}{8}}
\frac{\th_3(\tau,z)^2}{\eta(\tau)^3} 
~~~ (\mbox{for} ~\cC^{(\sNS)}_h).
\label{N=4 massive}
\end{eqnarray} 
 \item[massless representations $\cD^{(\sNS)}_{\ell}$] 
\begin{eqnarray}
\ch{\cN=4,(\sNS)}{0}(\ell=\frac 12;\tau,z) &=& 
q^{-1/8}\,
\sum_{n\in \bsz}\, \frac{1}{1+yq^{n-1/2}}\, q^{\frac{n^2}{2}}y^n
\frac{\th_3(\tau,z)}{\eta(\tau)^3} 
~~~ (\mbox{for} ~\cD^{(\sNS)}_{1/2}),
\label{N=4 massless matter}\\
\ch{\cN=4,(\sNS)}{0}(\ell=0;\tau,z) 
&=& 
q^{-1/8}\,
\sum_{n\in \bsz}\, \frac{(1-q)q^{\frac{n^2}{2}+n-\frac{1}{2}}y^{n+1}}
{(1+yq^{n+1/2})(1+yq^{n-1/2})} 
\frac{\th_3(\tau,z)}{\eta(\tau)^3} 
~~~ (\mbox{for}
~\cD^{(\sNS)}_{0}). \nn
&&
\label{N=4 grav}
\end{eqnarray}
\end{description}

The R-sector characters are obtained by the 1/2-spectral flow. Namely, 
\begin{eqnarray}
 && \ch{\cN=4,(\sR)}{}(h;\tau,z) = q^{\frac{1}{4}}y \, \ch{\cN=4,(\sNS)}{}
(h-\frac{1}{4};\tau,z+\frac{\tau}{2})~, ~~~ (\mbox{for}~ \cC^{(\sR)}_h)~, \nn
 && \ch{\cN=4,(\sR)}{0}(\ell;\tau,z) = q^{\frac{1}{4}}y \, \ch{\cN=4,(\sNS)}{0}
(\frac{1}{2}-\ell;\tau,z+\frac{\tau}{2})~, ~~~ (\mbox{for}~ \cD^{(\sR)}_{\ell})~. 
\label{N=4 R ch}
\end{eqnarray}

~


\section*{Appendix B: ~ Twisted Characters of $\cN=2$ and $\cN=4$ SCFTs}
\setcounter{equation}{0}
\def\theequation{B.\arabic{equation}}

~

In this appendix we summarize the definitions of the twisted characters of $\cN=2$ and $\cN=4$
superconformal algebras, according  to \cite{ES-G2orb,KawaiS2}.


~

\noindent
\underline{\bf $\cN=2$ twisted characters for the minimal model $M_k$}

We consider the characters of the $\cN=2$ SCA, twisted by 
the $\bz_2$-autormorphism
\begin{equation}
\sigma_L^{\cN=2} ~:~ T\,\longrightarrow\, T,~~~J\,\longrightarrow\, -J,~~~
G^{\pm}\,\longrightarrow\, G^{\mp} ,
\label{sigma twist}
\end{equation}
and express them as 
$\ch{(\al)}{[S,T]}$, 
where
$\al$ are the spin structures, and $S,T\in \bz_2$ signify the spatial 
and temporal boundary conditions associated with the $\sigma^{\cN=2}$-twist 
($S,T=1$ means twisted, and $S,T=0$ means untwisted).
We then have the following identities, 
\begin{eqnarray}
&& \ch{(\sNS)}{[0,1]}(\tau)= \ch{(\stNS)}{[0,1]}(\tau), ~~~
\ch{(\sNS)}{[1,0]}(\tau) = \ch{(\sR)}{[1,0]}(\tau), 
~~~ \ch{(\stNS)}{[1,1]}(\tau)= \ch{(\sR)}{[1,1]}(\tau) ~, 
\label{group 1} \\
&&  \ch{(\sR)}{[0,1]}(\tau)= \ch{(\stR)}{[0,1]}(\tau), ~~~
\ch{(\stNS)}{[1,0]}(\tau) = \ch{(\stR)}{[1,0]}(\tau), 
~~~ \ch{(\sNS)}{[1,1]}(\tau)= \ch{(\stR)}{[1,1]}(\tau) ,
\label{group 2}
\end{eqnarray}
and denote  
the twisted characters in the first line \eqn{group 1}
as `$\chi_{[0,1]}( \tau)$', `$\chi_{[1,0]}(\tau)$' and 
`$\chi_{[1,1]}(\tau)$' for brevity.
%
Especially, 
for the minimal models $M_{k}$, 
they are presented in \cite{ES-G2orb,KawaiS2}
(based on \cite{Dobrev,ZF2,Qiu,RY2}) as 
\begin{eqnarray}
\chi^k_{\ell \,[0,1]}(\tau)& =& 
\left\{
\begin{array}{ll}
\dsp  \frac{2}{\th_2(\tau)} \left(
\Th{2(\ell+1)}{4(k+2)}(\tau)+(-1)^k\Th{2(\ell+1)+4(k+2)}{4(k+2)}(\tau)
\right),   &  ~~ (\ell~:~\mbox{even}) ,\\
0, & ~~(\ell~:~\mbox{odd}).
\end{array}
\right.
\nn
\chi^k_{\ell\,[1,0]}(\tau)&=& \frac{1}{\th_4(\tau)}
\,\left(\Th{\ell+1-\frac{k+2}{2}}{k+2}(\tau)-
\Th{-(\ell+1)-\frac{k+2}{2}}{k+2}(\tau)\right)  \nn
&= & \frac{1}{\th_4(\tau)}\left(\Th{2(\ell+1)-(k+2)}{4(k+2)}(\tau)
 + \Th{2(\ell+1)+3(k+2)}{4(k+2)}(\tau) \right. \nn
 && \left. - \Th{-2(\ell+1)-(k+2)}{4(k+2)}(\tau)
 - \Th{-2(\ell+1)+3(k+2)}{4(k+2)}(\tau) \right),
\nn
\chi^k_{\ell\,[1,1]}(\tau)&=& 
\frac{1}{\th_3(\tau)}
\left(\Th{2(\ell+1)-(k+2)}{4(k+2)}(\tau)
 +(-1)^k \Th{2(\ell+1)+3(k+2)}{4(k+2)}(\tau) \right. \nonumber \\
 && \left. +(-1)^{\ell} \Th{-2(\ell+1)-(k+2)}{4(k+2)}(\tau)
 +(-1)^{k+\ell} \Th{-2(\ell+1)+3(k+2)}{4(k+2)}(\tau) \right).
\label{twisted minimal}
\end{eqnarray}
The conformal weights of the ground states corresponding to the first characters are
\begin{eqnarray}
h= 
h_{\ell} \equiv 
\frac{\ell(\ell+2)}{4(k+2)},
\end{eqnarray}
while those for the second and third ones are given by
\begin{eqnarray}
h= 
h_{\ell}^{t} \equiv  
\frac{k-2+(k-2\ell)^2}{16(k+2)}+\frac{1}{16}.
\label{h t}
\end{eqnarray}
Note that only the states with the vanishing $U(1)$-charges can contributes to the relevant characters. 
Note also that 
$\chi^k_{k-\ell\,[1,0]}=\chi^k_{\ell\,[1,0]}$, 
$\chi^k_{k-\ell\,[1,1]}=\chi^k_{\ell\,[1,1]}$.
Due to these relations the corresponding fields are identified,
leaving only $\ell=0, 1,\ldots, \left\lb \frac{k}{2}\right\rb$ as 
independent primary fields.

The modular transformations of the twisted $\cN =2$ characters are 
\begin{eqnarray}
&&\hskip-15mm
\chi^k_{\ell\,[0,1]}(\tau+1)= 
e^{2\pi i \left(h_\ell -\frac{k}{8(k+2)}\right)}\,
\chi^k_{\ell\,[0,1]}(\tau), 
\hskip5mm
\chi^k_{\ell\,[0,1]}\left(-\frac{1}{\tau}\right)
=\sum_{\ell'=0}^k\, (-1)^{\ell/2} 
S_{\ell,\ell'}\,
\chi^k_{\ell'\,[1,0]}(\tau),  \nn
&&\hskip-15mm
\chi^k_{\ell\,[1,0]}(\tau+1)=
e^{2\pi i\left(h^t_\ell-\frac{k}{8(k+2)}\right)}\, 
\chi^k_{\ell\,[1,1]}(\tau), 
\hskip5mm
\chi^k_{\ell\,[1,0]}\left(-\frac{1}{\tau}\right)=
\sum_{\ell'=0}^k\, S_{\ell,\ell'}(-1)^{\ell'/2}\, 
\chi^k_{\ell'\,[0,1]}(\tau), \nn
&&\hskip-15mm \chi_{\ell\,[1,1]}(\tau+1)= 
e^{2\pi i \left(h^t_\ell -\frac{k}{8(k+2)}\right)}\,
\chi^k_{\ell\,[1,0]}(\tau)  , 
\hskip5mm
\chi^k_{\ell\,[1,1]}\left(-\frac{1}{\tau}\right)=
\sum_{\ell'=0}^k\, \widehat{S}_{\ell,\ell'}\,
\chi^k_{\ell'\,[1,1]}(\tau)  .
\label{modular twisted minimal} 
\end{eqnarray}
Here 
$ S_{\ell,\ell'}\equiv \sqrt{\frac{2}{k+2}}
\sin\left(\frac{\pi(\ell+1)(\ell'+1)}{k+2}\right)$ is 
the modular S-matrix of the $SU(2)$ WZW model at level $k$, 
and 
$\widehat{S}_{\ell,\ell'}\equiv e^{\frac{\pi i}{2}\left(\ell+\ell'-\frac{k}{2}\right)}\, 
S_{\ell,\ell'}$.


Let us briefly comment on 
the remaining minimal model characters appearing in the second line
\eqn{group 2}.  
For example, for the  $[0,1]$-type boundary condition in the R-sector,
almost all the characters vanish,
except for the special representation 
generated by the non-degenerate Ramond ground state with $ h=\frac{\hat{c}}{8}$, $Q=0$, that is, 
$
\ell = \frac{k}{2},
$ 
$
m= \pm (\frac{k}{2}+1)
$
with 
$
k\in 2\bz_{>0}.
$
The corresponding character equal $\pm 1$, where the sign ambiguity is 
just due to the action of 
$\sigma^{\cN=2}_L$ on primary states.

~


\noindent
\underline{\bf $\cN=4$ twisted characters}


We next summarize the twisted $\cN=4$ characters defined by the $\sigma^{(1)}_L$ 
and $\sigma^{(3)}_L$-twists
in the unitary irrep.'s 
 of the $\cN=4$ SCA with $c=6$.
We first focus on the $\sigma^{(3)}_{L}$-twist for the boundary conditions given in \eqn{group 1}. 
The key formula is the spectral flow decomposition of the $\cN=4$ characters by the $\cN=2$ ones \cite{ET}, 
written schematically as 
\begin{align}
\ch{\cN=4,(\sNS)}{}(*;\tau,z) 
& = \sum_{n\in \bz}  q^{n^2}y^{2n}
\ch{\cN=2,(\sNS)}{}(*;\tau,z+n\tau),
\label{N=4 N=2 ch}
\end{align}
for the NS-sector, 
where $n \in \bz$ is identified with 
the $n$-th spectral flow sector. 
It is again the simplest to evaluate the case of $[S,T]=[0,1]$ ({\em i.e.} with the insertion 
of $\sigma^{(3)}_{L}$ into the trace).
This just yields an extra phase factor $(-1)^n$ in each $n$-th spectral flow sector 
in the decomposition \eqn{N=4 N=2 ch}, and we obtain the desired character formulas (by setting $z=0$): 
\begin{description}
 \item[massive representation $\cC^{(\sNS)}_h$] 
\begin{align}
\tr_{\cC^{(\sNS)}_h}\lb\sigma^{(3)}_{L}q^{L_0-\frac{1}{4}} \rb
&= q^{h-\frac{1}{8}} \sum_{n\in\bz}
 (-1)^n q^{\frac{n^2}{2}}\frac{\th_3(\tau)}{\eta(\tau)^3}
= q^{h-\frac{1}{8}}
\frac{\th_3(\tau)\th_4(\tau)}{\eta(\tau)^3} 
\nn
&
\equiv \frac{2q^{h-\frac{1}{8}}}{\th_2(\tau)} = : \chi_{[0,1]}(h; \tau).
\label{twisted N=4 massive} 
\end{align} 
\item[massless representations $\cD^{(\sNS)}_{\ell}$] 
\begin{align} 
\tr_{\cD_{1/2}^{(\sNS)}}\lb\sigma^{(3)}_{L}q^{L_0-\frac{1}{4}} \rb
&= 
q^{-1/8}\,
\sum_{n\in \bsz}\, (-1)^{n+1}\frac{1}{1+q^{n-1/2}}\, 
\frac{\th_3(\tau)}{\eta(\tau)^3} \equiv 0,
\label{twisted N=4 massless} \\
\tr_{\cD_{0}^{(\sNS)}}\lb\sigma^{(3)}_{L}q^{L_0-\frac{1}{4}} \rb
& =
q^{-1/8}\,
\sum_{n\in \bsz}\,  (-1)^n
 \frac{(1-q)q^{\frac{n^2}{2}+n-\frac{1}{2}}}
{(1+q^{n+1/2})(1+q^{n-1/2})} 
\frac{\th_3(\tau)}{\eta(\tau)^3}
\nn
&=  q^{-1/8} 
\frac{\th_3(\tau)\th_4(\tau)}{\eta(\tau)^3}
\equiv \chi_{[0,1]}(h=0;\tau,z).
\label{twisted N=4 grav} 
\end{align}
The second line of \eqn{twisted N=4 grav} follows from the identity
\begin{eqnarray}
 && \frac{(1-q)q^{n-\frac{1}{2}}}
{(1+q^{n+1/2})(1+q^{n-1/2})} 
= 1- \frac{1}{1+q^{n-\frac{1}{2}}} -
\frac{q^{n+\frac{1}{2}}}{1+q^{n+\frac{1}{2}}}.
\label{identity 1}
\end{eqnarray}
\end{description}


We next consider the $\sigma_L^{(1)}$-twist. 
Since the $\sigma^{(1)}_{L}$-twist acts as $J(\equiv 2J^3)\,\rightarrow\, -J$
on the $U(1)_R$-current of the underlying $\cN=2$ SCA, 
the spectral flow sectors of $n\neq 0$ cannot contribute 
when $\sigma^{(1)}_{L}$ is inserted into the trace. 
Thus, the wanted characters 
should be equal to the ones for the $\cN=2$ non-degenerate representations, that is, 
\begin{align}
\tr_{\cC^{(\sNS)}_h}\lb\sigma^{(1)}_{L}q^{L_0-\frac{1}{4}} \rb
&= \frac{q^{h-1/8}}{\eta(\tau)} \cdot
\sqrt{\frac{2\eta(\tau)}{\th_2(\tau)}}\cdot
\sqrt{\frac{\th_3(\tau)\th_4(\tau)}{\eta(\tau)^2}} 
\equiv 
\chi_{[0,1]}(h; \tau),
\label{twisted N=4 massive 2} 
\\
\tr_{\cD_{1/2}^{(\sNS)}}\lb\sigma^{(1)}_{L}q^{L_0-\frac{1}{4}} \rb
&= 0,
\label{twisted N=4 massless 2} \\
\tr_{\cD_{0}^{(\sNS)}}\lb\sigma^{(1)}_{L}q^{L_0-\frac{1}{4}} \rb
& =
\frac{q^{-1/8}}{\eta(\tau)} \cdot
\sqrt{\frac{2\eta(\tau)}{\th_2(\tau)}}\cdot
\sqrt{\frac{\th_3(\tau)\th_4(\tau)}{\eta(\tau)^2}} 
\equiv 
\chi_{[0,1]}(h=0; \tau).
\label{twisted N=4 grav 2} 
\end{align}
They indeed coincide with those of $\sigma^{(3)}_L$-twisting \eqn{twisted N=4 massive}, 
\eqn{twisted N=4 massless} and \eqn{twisted N=4 grav}.%
\footnote
   {This coincidence would be anticipated. However, it is not necessarily self-evident because 
 the automorphisms $\sigma^{(1)}_L$ and $\sigma^{(3)}_L$ are interpolated only 
 by an {\em outer}-automorphism of the $\cN=4$ SCA, as opposed to the case of 
{\em e.g.} $\widehat{SU}(2)_k$.}
The $\sigma^{(2)}_L$-twisting leads to the same formulas, too.


The character formulas for other boundary conditions are just 
determined by the modular transformations.
We denote the spin structures as well as the boundary conditions of $\sigma^{(\alpha)}_{L}$  
such as $\{\NS,\, [S,T]\}$. 
Starting from the character formula of $\{\NS,\,[0,1]\}$ given above, 
we find that there are three types of non-trivial characters 
$\chi_{[0,1]}(h;\tau)$, $\chi_{[1,0]}(h;\tau)$, $\chi_{[1,1]}(h;\tau)$;
\begin{eqnarray}
&&  \{\NS,\, [0,1]\}, ~  \{\tNS,\, [0,1]\}~:~~  
\chi_{[0,1]}(h;\tau) \equiv \frac{2 q^{\frac{p^2}{2}}}{\th_2(\tau)}~, ~~~ 
(h=\frac{p^2}{2}+\frac{1}{8}), \nn
&&
 \{\NS,\, [1,0]\}, ~  \{\R,\, [1,0]\} ~:~~
\chi_{[1,0]}(h;\tau) \equiv \frac{2 q^{\frac{p^2}{2}}}{\th_4(\tau)}~, ~~~ 
(h=\frac{p^2}{2}+\frac{1}{4}), \nn
&&
\{\tNS,\, [1,1]\}, ~  \{\R,\, [1,1]\} ~:~~  
\chi_{[1,1]}(h;\tau) \equiv \frac{2 q^{\frac{p^2}{2}}}{\th_3(\tau)}~, ~~~ 
(h=\frac{p^2}{2}+\frac{1}{4}).
\label{twisted N=4 characters}
\end{eqnarray}

There still remain the boundary conditions presented in \eqn{group 2}.
We briefly describe them although only the ones listed in \eqn{twisted N=4 characters} are necessary 
in the main text,
\begin{eqnarray}
&& \tr_{\cC^{(\sR)}_h} \left[ \sigma^{(\al)}_L q^{L_0-\frac{1}{4}}\right]
 =\tr_{\cD^{(\sR)}_{1/2}} \left[ \sigma^{(\al)}_L
			   q^{L_0-\frac{1}{4}}\right]
=0~, ~~~ \tr_{\cD^{(\sR)}_{0}} \left[ \sigma^{(\al)}_L 
q^{L_0-\frac{1}{4}}\right] = \pm 1, 
\label{twisted ch R 0 1}
\end{eqnarray}
$(\any \alpha = 1,2,3)$.
The sign ambiguity in the formula for $\cD^{(\sR)}_0$ is 
due to the same reason as above.
We also obtain the same results for the $\{\tR,\,(0,1)\}$-characters.
It is trivial to modular transform these results to obtain the remaining ones
$\{\tNS,\,[1,0]\}$, $\{\NS,\,[1,1]\}$ ($\{\tR,\,[1,0]\}$, $\{\tR,\,[1,1]\}$).


\end{document}